\documentclass[conference,10pt]{IEEEtran}

\usepackage{tikz}
\usepackage{amsmath}
\usepackage{amsthm}
\usepackage{filecontents}
\usepackage{caption}
\usepackage{subcaption}
\usepackage{bbm}
\usepackage{bm}
\usepackage{algcompatible}
\usepackage{algorithm}
\usepackage{enumitem}
\usepackage{multicol}
\usepackage{algpseudocode} 
\usepackage{booktabs}
\usepackage{arcs}
\usepackage[nowidering]{yhmath}
\usepackage[normalem]{ulem}
\usepackage{epstopdf}
\usepackage{pifont}
\usepackage{epsfig,endnotes}
\usepackage{graphicx}
\usepackage{color}
\usepackage{textcomp}
\usepackage{multirow}
\usepackage{booktabs}
\usepackage{fontawesome}
\usepackage{xcolor}
\usepackage{algpseudocode}
\usepackage{makecell}
\usepackage{balance}
\usepackage{filecontents}

\usepackage{xspace}
\usepackage{caption}
\usepackage{fancyhdr}
\usepackage{setspace}
\usepackage{listings}
\usepackage{environ}
\usepackage{hyperref}

\newcommand{\mcl}{\mathcal}

\newcommand{\mbb}{\mathbb}

\def \RR {{\mathbb{R}}}

\newtheorem{theorem}{Theorem}
\newtheorem{definition}{Definition}
\newtheorem{lemma}{Lemma}

\newtheorem{assumption}{Assumption}

\def\BW#1{{\textcolor{violet}{#1}}}
\def\YF#1{{\textcolor{orange}{#1}}}

\def\xs#1{\textcolor{blue}{#1}}
\def\todo#1{\textcolor{orange}{#1}}

\def\vv{{\bm{v}}}
\def\vw{{\bm{w}}}
\def\vx{{\bm{x}}}

\def \mI {{\bm{I}}}

\def \mM {{\bm{M}}}

\def \RR {{\mathbb{R}}}

\def \RE {}

\usepackage{verbatim}
\usepackage{cleveref}

\IEEEoverridecommandlockouts\IEEEpubid{\makebox[\columnwidth]{ 979-8-3503-5171-2/24/\$31.00 $\copyright$2024 IEEE \hfill}\hspace{\columnsep}\makebox[\columnwidth]{ }}
\begin{document}

\title{Towards Practical Overlay Networks for Decentralized Federated Learning 
}

\author{Yifan Hua\textsuperscript{1}, Jinlong Pang\textsuperscript{1}, Xiaoxue Zhang\textsuperscript{1}, Yi Liu\textsuperscript{1}, Xiaofeng Shi\textsuperscript{1}, Bao Wang\textsuperscript{2}, Yang Liu\textsuperscript{1}, Chen Qian\textsuperscript{1}\\
\textit{\textsuperscript{1}University of California Santa Cruz, \textsuperscript{2}University of Utah}\\
\{yhua294, pang14, xzhan330, yliu634, xshi24, yangliu, cqian12\}@ucsc.edu  \space\space   bwang@math.utah.edu
}

\maketitle
\thispagestyle{plain}
\pagestyle{plain}
\begin{abstract}
Decentralized federated learning (DFL) uses peer-to-peer communication to avoid the single point of failure problem in federated learning and has been considered an attractive solution for machine learning tasks on distributed devices. We provide the first solution to a fundamental network problem of DFL: what overlay network should DFL use to achieve fast training of highly accurate models, low communication, and  decentralized construction and maintenance? Overlay topologies of DFL have been investigated, but 
no existing DFL topology includes decentralized protocols for network construction and topology maintenance. Without these protocols, DFL cannot run in practice. This work presents an overlay network, called FedLay, which provides fast training and low communication cost for practical DFL. FedLay is the first solution for constructing near-random regular topologies in a decentralized manner and maintaining the topologies under node joins and failures. Experiments based on prototype implementation and simulations show that FedLay achieves the fastest model convergence and highest accuracy on real datasets compared to existing DFL solutions while incurring small communication costs and being resilient to node joins and failures. 

\end{abstract}

\section{Introduction}

Training machine learning (ML) models using data collected by distributed devices,
such as mobile and IoT devices, is crucial for modern ML. 
Federated learning (FL) \cite{pmlr-v54-mcmahan17a,karimireddy2020scaffold,li2020federated,pathak2020fedsplit,reddi2020adaptive,zhou2021truthful, acar2021federated} has become a popular ML paradigm that allows a large number of clients (end systems, edge nodes, etc.) to train 
ML models collaboratively without directly sharing training data.
FL uses a central server or cloud to orchestrate clients for training ML models and iterates the following procedure: The 
server creates a global model by aggregating the local ML models collected from the clients and then sends it to clients for edge applications; the ML models are updated at the clients. 

\begin{figure}
    \centering
    \includegraphics[width=0.45\textwidth]{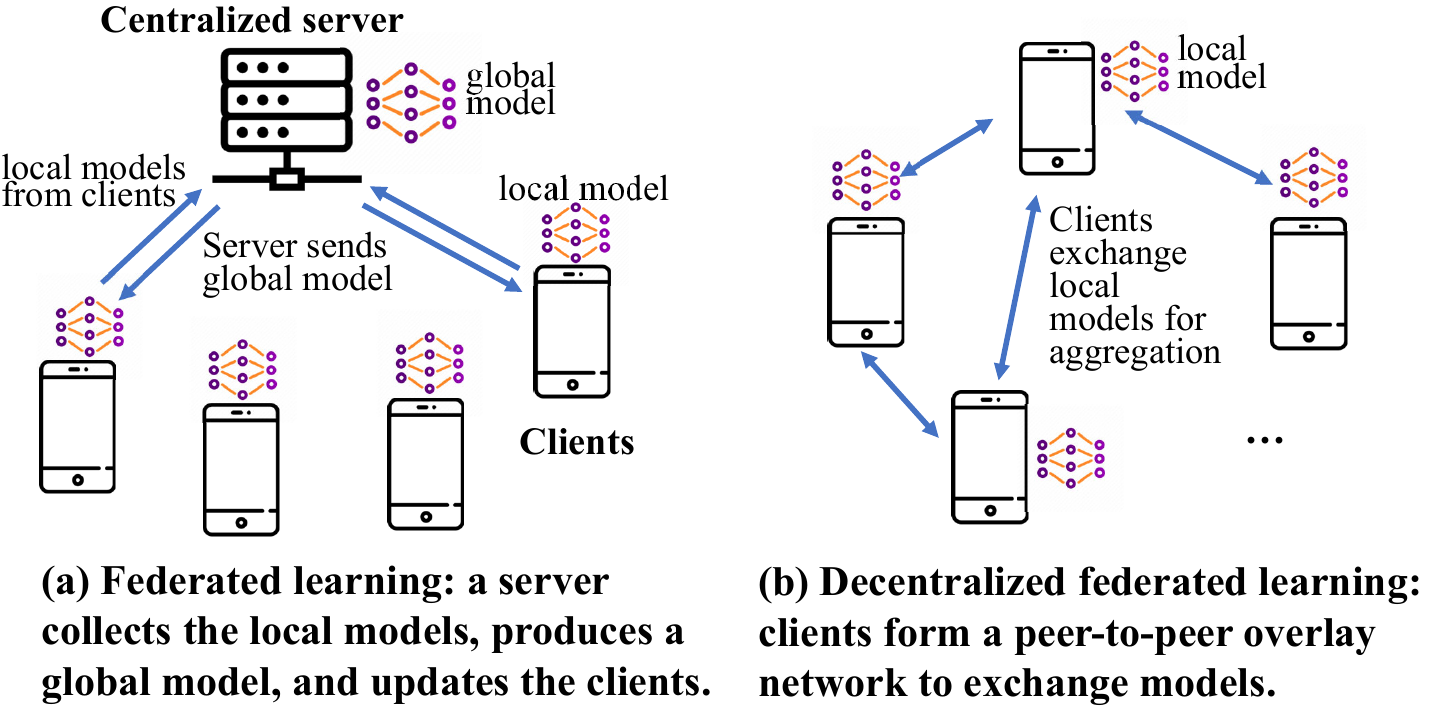}
     \caption{Federated learning v.s. decentralized federated learning}
    \label{fig:FLDFL}
    \vspace{-1ex}
\end{figure}

An abstraction of FL is shown in Fig.~\ref{fig:FLDFL}(a). 
Compared to collecting raw data from distributed devices and performing centralized ML, FL has several main advantages, including saving communication costs on limited-bandwidth devices, preserving data privacy, and being compatible with country or organization regulations that prohibit direct data sharing. 


However, the drawbacks of FL are also prominent and have been studied and widely mentioned in the literature \cite{li2020federated,he2018cola,wang2021edge}. For example, the central orchestration server that frequently exchanges models with clients clearly presents a bottleneck and becomes a typical \textit{single point of failure} \cite{li2020federated,Beltran2022DFL}. In addition, the server is also a single point of attack: adversaries can make all clients use tampered ML models by attacking the server. There is even a risk that the server itself is malicious, which might distribute incorrect global models or collect sensitive information from the 
clients. 

\begin{table*}[thbp]
	\centering
	\resizebox{\textwidth}{!}{
		\renewcommand\arraystretch{.5}
		\setlength{\tabcolsep}{2mm}
		\begin{tabular}{cccccc}
			\toprule
			Overlay network & Decentralized construction& Node degree & Model convergence & Resilience to churn& Other comments\\
			\midrule
			Ring \cite{he2018cola}& Not discussed & 2& Slow & Not discussed & \\
			\midrule
			2D grid \cite{he2018cola}& Not discussed & 4& Slow & Not discussed & \\
			\midrule
			Complete graph \cite{he2018cola} & Not discussed & $N$-1& Fast & Not discussed & \\
			\midrule 
            Dynamic chain \cite{elgabli2020communication}& Not discussed & 2& Faster than ring & Not discussed & \\
			\midrule
            D-Cliques \cite{bellet2022d} & Unknown & $|C|$-1 & Fast & Not discussed &  Assume global knowledge\\
            \midrule
            Clustering \cite{al-abiad2022cluster}& Not discussed & $|C|$-1 & Fast & Not discussed & Bottlenecks exist \\
			\midrule
            Hypercube \cite{vogels2022beyond} & Not discussed & $O(\log N)$ & Fast & Not discussed &  \\
            \midrule
            Torus \cite{vogels2022beyond} & Not discussed & $d$ & Fast & Not discussed &  \\
            \midrule
            Ramanujan \cite{chow2016expander,hua2022efficient} & Not discussed & $d$ & Fast & Not discussed &  \\
			\midrule
            Random $d$-graph \cite{chow2016expander,hua2022efficient} & Unknown & $d$ & Fast & Not discussed & Assume global knowledge \\
			\midrule
            \textbf{FedLay (this work)}& \textbf{Yes} & \textbf{$d$} & \textbf{Fast}& \textbf{Yes} &\textbf{Address device/data heterogeneity}\\ 
			\bottomrule
	\end{tabular}}
	\caption{List of overlay network topologies for DFL. $N$ is the 
 number of clients. $d$ is a small constant for node degree, usually around 10. $|C|$ is the size of each cluster/clique, usually bigger than values of $d$. Other than FedLay, only D-Cliques and \cite{hua2022efficient} discuss its construction algorithm but still assumes global knowledge.}
        \vspace{-2ex}
	\label{table:com}
 
\end{table*}

Decentralized federated learning (DFL) emerged recently \cite{he2018cola,sun2021decentralized,Beltran2022DFL} to resolve the above problems of FL, by removing the involvement of the central server. 
As shown in  Fig.~\ref{fig:FLDFL}(b), in a DFL system, 
DFL clients form a peer-to-peer (P2P) network and keep exchanging their models using P2P communication. In most cases, the data on different clients are not identically and independently distributed (non-iid) \cite{he2018cola,sun2021decentralized,Beltran2022DFL}, hence  the trained local ML models are substantially different from each other.
After sufficient model exchanges,
the local models on the clients may converge to a model that correctly reflects the features of data from all clients.  

This work focuses on \textit{a fundamental network problem} of DFL: what overlay network is ideal for DFL in practice? An overlay network of DFL is a logical network on top of the physical networks. It specifies which pairs of clients should exchange their local models: two clients exchange models if they are overlay neighbors. 
An ideal overlay network for DFL needs to satisfy a few requirements \cite{Beltran2022DFL} including 1) a decentralized construction protocol that can build the overlay topology;  2) fast convergence of local models to high accuracy; 3) small node degree that can maintain low bandwidth cost on clients for exchanging models with a limited number of neighbors;
4) resilient to client dynamics such as client joins, leaves, and failures -- they are also called as \textit{churn}. 

Table \ref{table:com} shows a list of overlay network topologies that have been studied for DFL. We find that most of these existing studies do not pay attention to whether the proposed topologies  can be constructed by decentralized protocols and resilient to churn. These two requirements are common networking/distributed system problems and might not be the focus for ML researchers. 

\textbf{DFL cannot work as a practical system without a decentralized construction protocol for its overlay network.} 
For example, recent work suggests that Ramanujan graphs provide fast convergence and accurate models for DFL \cite{sun2021stability,hua2022efficient}. However, the decentralized construction of Ramanujan graphs is unknown. 
Centralized construction/maintenance contradicts the main purpose of DFL: avoid the single point of failure/attack.

We propose a fully decentralized overlay network for DFL, called FedLay, which achieves all four requirements discussed above, namely decentralized construction, fast convergence to accurate models, small communication cost,  and resilience to churn. FedLay does not need a centralized server at any stage and all clients run the same suite of distributed protocols. 
The FedLay protocol suite includes two sets of protocols: 1) a set of \textit{Neighbor Discovery and Maintenance Protocols (NDMP)} to build the overlay network and recover it from churn; and 2) a \textit{Model Exchange Protocol (MEP)} to achieve fast model convergence for heterogeneous clients and asynchronous communication.  
The FedLay topology is motivated by the near-random regular topologies that have been proposed for data center networks \cite{Jellyfish,S2-ICNP}. \textbf{However, \cite{Jellyfish} \cite{S2-ICNP} are centralized protocols for data centers and cannot be applied to DFL.} 
To our knowledge, FedLay is the first solution for constructing near-random regular topologies in a decentralized manner and maintaining the topologies under node joins and failures.
FedLay also considers other practical issues, including non-iid data and asynchronous communication with heterogeneous clients. 

The contributions of this work are summarized as follows. 
\begin{itemize}[leftmargin=*]
\vspace{-.5ex}
    \item We identify three topology metrics related to DFL convergence and evaluate various overlay topologies of DFL. We find that FedLay outperforms all other topologies. 
    FedLay, as a decentralized network, has almost identical results on all three metrics to the best result among the 100 randomly generated regular graphs (in a centralized way).
    \item We design and implement the FedLay protocol suite. \textbf{To our knowledge, FedLay is the first DFL overlay network that provides decentralized protocols for construction, churn recovery, and model aggregation.}
    \item We evaluate FedLay using both prototype implementation and simulations on real ML datasets. We find that FedLay achieves the highest average model accuracy and fastest convergence compared to other DFL methods. It also has 
    small communication cost and strong resilience to churn. 
    \vspace{-.5ex}
\end{itemize} 


The rest of this paper is organized as follows. Section \ref{sec:topology} presents the metrics for selecting DFL overlay topologies and the details of the topology of FedLay. The design of FedLay protocol suite, including the topology construction and maintenance protocols and model aggregation protocol, is presented in Section \ref{sec:design}. Section \ref{sec:evaluation} shows the evaluation results of FedLay as well as existing DFL overlay networks  on real ML datasets. 
We present related work in Section \ref{sec:related} and conclude this work in Section \ref{sec:conclusion}.

\section{Overlay Topology of FedLay}
\label{sec:topology}
This section presents the topology of FedLay and the intuition behind using this topology. 
We first explore what topology metrics can be used to evaluate the convergence speed under small node degrees. 
Then we design the FebLay topology and use numerical results to show its advantages.
The decentralized construction and maintenance under churn will be presented in the next section.

\vspace{-1ex}
\subsection{Assumptions}
\vspace{-1ex}
\label{sec:assumption}
This work is based on the following assumptions: All the devices in FedLay are already connected to the Internet where they can directly access each other using TCP/IP. All clients train the same neural network models. Clients are honest and benign. The security problems under dishonest clients will be considered in future work. 

\vspace{-2ex}
\subsection{Three metrics for DFL topologies}
\label{sec:metrics}
\vspace{-1ex}
A DFL topology can be modeled as an undirected graph $G=(V,E)$, where each node $v\in V$ represents
a client in the DFL system and each link $e=(u,v) \in E$ indicates that two clients $u$ and $v$ 
will exchange local ML models -- $u$ and $v$ are thus called neighbors. 
We assume clients have equal roles in the overlay and similar numbers of neighbors.

\subsubsection{Expander property and convergence factor.}

An important notion in DFL, or general decentralized optimization algorithms, is the mixing matrix $\mM$ of the graph $G$. The $i$-th row of $\mM$ denotes the weights used for aggregating local models of the neighboring nodes to update the model of the $i$-th client. 
Hence the adjacency matrix of an overlay network and its Metropolis-Hastings matrix are both mixing matrices \cite{boyd2004fastest}. The symmetric property of ${\mM}$ indicates that its eigenvalues are real and can be sorted in non-increasing order. Let $\lambda_i({\mM})$ denote the $i$-th largest eigenvalue of ${\mM}$, then we have  $\lambda_1({\mM})=1>\lambda_2({\mM})\geq \cdots \geq \lambda_N({\mM})>-1$ based on the spectral property of the mixing matrix~\cite{boyd2004fastest}. The constant $\lambda=\lambda({\mM}):=\max\{|\lambda_2({\mM})|,|\lambda_N({\mM})|\}$ has been used to characterize optimization error (a measure of training loss) and generalization gap (a measure of test accuracy) of DFL. In particular, it is shown that the optimization error and generalization gap -- for a typical DFL framework, Decentralized Federated Averaging (DFedAvg) -- are bounded, respectively, by $\mathcal{O}\Big(\frac{1}{(1-\lambda)^2}\Big)$ and $\mathcal{O}\Big(2\lambda^2 +  4\lambda^2 \ln \frac{1}{\lambda} + 2\lambda + \frac{2}{\ln \frac{1}{\lambda}}\Big)$ -- in terms of $\lambda$ \cite{sun2021decentralized,hua2022efficient}. Notice that both $\frac{1}{(1-\lambda)^2}$ and $2\lambda^2 +  4\lambda^2 \ln \frac{1}{\lambda} + 2\lambda + \frac{2}{\ln \frac{1}{\lambda}}$ are increasing functions of $\lambda\in (0,1)$.

Per the above discussion, to achieve good convergence and generalization, a topology needs to have a $\lambda$ sufficiently smaller than 1 and hence achieve a small value of $\frac{1}{(1-\lambda)^2}$ and 
$2\lambda^2 +  4\lambda^2 \ln \frac{1}{\lambda} + 2\lambda + \frac{2}{\ln \frac{1}{\lambda}}$. 
Thus we define the first topology metric, called the \textit{convergence factor} of $G$: $c_G=\frac{1}{(1-\lambda)^2}$.

Note that when $\frac{1}{(1-\lambda)^2}$ is minimized, $2\lambda^2 +  4\lambda^2 \ln \frac{1}{\lambda} + 2\lambda + \frac{2}{\ln \frac{1}{\lambda}}$ is also minimized. Hence for the sake of simplicity, we do not need another factor.

\subsubsection{Network diameter.}
The diameter of a network is the longest length of all shortest paths calculated in the network. It reflects the network distance between the two most distant nodes. The intuition of considering this metric is that the network diameter can represent the maximum latency that the local model trained on the data of a client can propagate to all clients in the network. 

\subsubsection{Average length of shortest paths.}
The third metric is the average length of all shortest paths in the network. The intuition of considering this metric is that the average length can represent the average latency that a local model can propagate to a random client. 


\vspace{-1ex}
\subsection{FedLay topology}
\label{sec:fedlaytopology}

The FedLay topology is motivated by the research on near-random regular graphs for data center networks \cite{Jellyfish,S2-ICNP,S2-TPDS} and DFL \cite{chow2016expander,hua2022efficient}.
Recent theoretical studies show that Ramanujan graphs can provide small values of the spectral expander property $\lambda$ and hence achieve `optimal' convergence with a constant node degree $d$ \cite{hua2022efficient}. 
However, a large Ramanujan graph 
cannot be generated even by centralized construction. 
Hence, random regular graphs (RRGs) can be used instead, which are approximately Ramanujan for a large network size $n$ \cite{chow2016expander}. 
In addition, prior research on data center networks \cite{Jellyfish} also shows that near-RRGs achieve the smallest average length of shortest paths among known graphs with a fixed node degree $d$. Note that RRGs cannot be generated with any deterministic algorithm either. Hence, near-RRGs are usually used in practice, which are considered close enough to RRGs \cite{Jellyfish,S2-ICNP}. 
near-RRGs can achieve ideal values on both the convergence factor and shortest path lengths.  

\begin{figure}[t]
    \centering
    \includegraphics[width=0.48\textwidth]{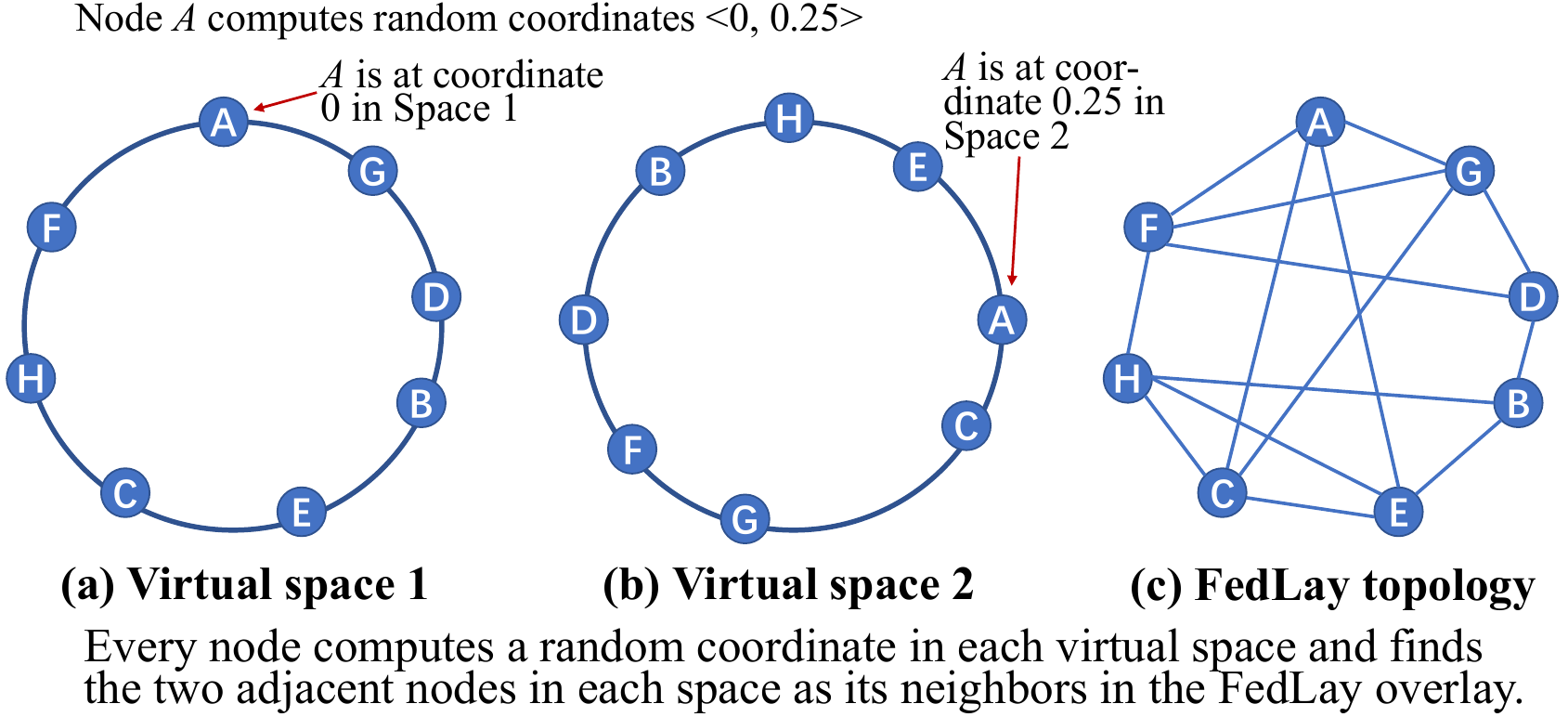}
     \caption{An example of FedLay topology}
    \label{fig:topo}
    \vspace{-2ex}
\end{figure}

A practical problem is that all existing near-RRGs are constructed by centralized methods \cite{Jellyfish,S2-ICNP,S2-TPDS,hua2022efficient}. The main challenge of a decentralized construction of near-RRGs
is to allow a node to select a neighbor from all other nodes with equal likelihood while this node does not know the entire network. 
We propose to use a virtual coordinate system to solve this problem, inspired by \cite{S2-ICNP}. \textbf{Note that \cite{S2-ICNP} is a centralized protocol and our main contribution is a decentralized construction for DFL.}


In FedLay, each node computes a set of \textit{virtual} coordinates $C$, which is 
an $L$-dimensional vector $<x_1,x_2,...,x_L>$ where each
element $x_l$ is a random real number in range $[0, 1)$. In practice, $x_i$ can be computed as $H(IP_x|i)$ where $H$ is a publicly known hash function and  $IP_x$ is $x$'s IP address. 

We define $L$ virtual ring spaces. 
In the $i$-th ring space, a node
is \textit{virtually} placed on a ring based on the value of its $i$-th coordinate $x_i$. 
The coordinates of each space are circular, with 0 and 1 being superposed at the top-most point of the ring and 0.5 being the bottom-most point. 
If the coordinates of two different nodes are identical in one  space, their orders on the ring are determined by the values of their IP addresses. For ease of presentation, we assume all coordinates on a ring are different. As shown in the example in Fig.~\ref{fig:topo}, there are 8 nodes and each of them computes a set of two-dimensional \textit{random} coordinates $<x_1,x_2>$. There are two virtual ring space as shown in Figs.~\ref{fig:topo} (a) and (b) and every node is on a position of the $i$-th ring based on its random coordinate $x_i$. Note, all spaces are virtual and they have no relationship to the geographic locations of the nodes. 

In each virtual space, every node $u$ has two adjacent nodes on the ring, based on the order of their coordinate values. 
$u$ will find the adjacent nodes from all spaces as its overlay neighbors  (by a decentralized protocol described later) for model exchange. 
In the example of Fig.~\ref{fig:topo}(c), every node finds its adjacent nodes in two spaces and form the FedLay overlay. So most nodes have four neighbors in the overlay but there are a few ones, like node $B$, has only three neighbors because $D$ is adjacent to $B$ on both rings.  
Hence, for $L$ spaces, every node has at most $2L$ neighbors. 
$L$ can then be considered as a parameter for communication and convergence trade-off: with a bigger $L$, nodes have more neighbors for model exchanges but increased communication cost.  

\begin{figure*}[t]
    \centering
    \begin{subfigure}[b]{0.30\textwidth}
        \centering
        \includegraphics[width=\textwidth]{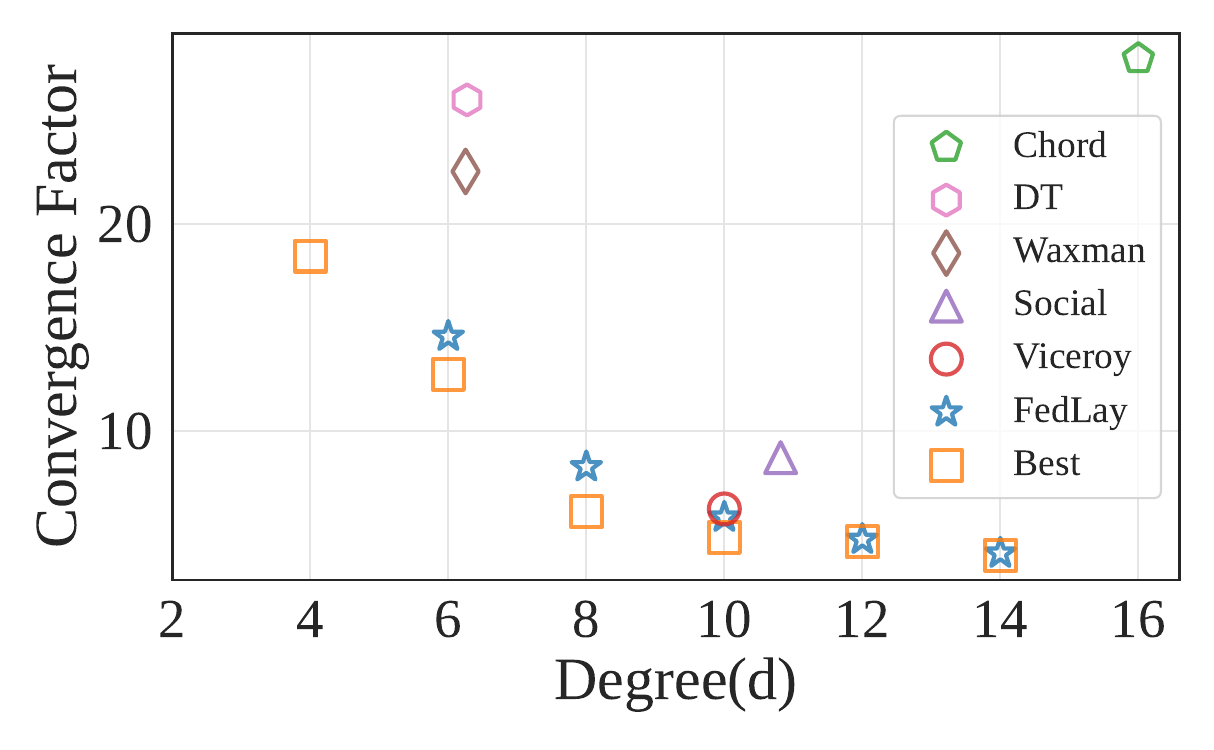}
        \vspace{-5ex}
        \caption{Convergence factor}
        \label{fig:expander1-a}
    \end{subfigure}
    \begin{subfigure}[b]{0.308\textwidth}
        \centering
        \includegraphics[width=\textwidth]{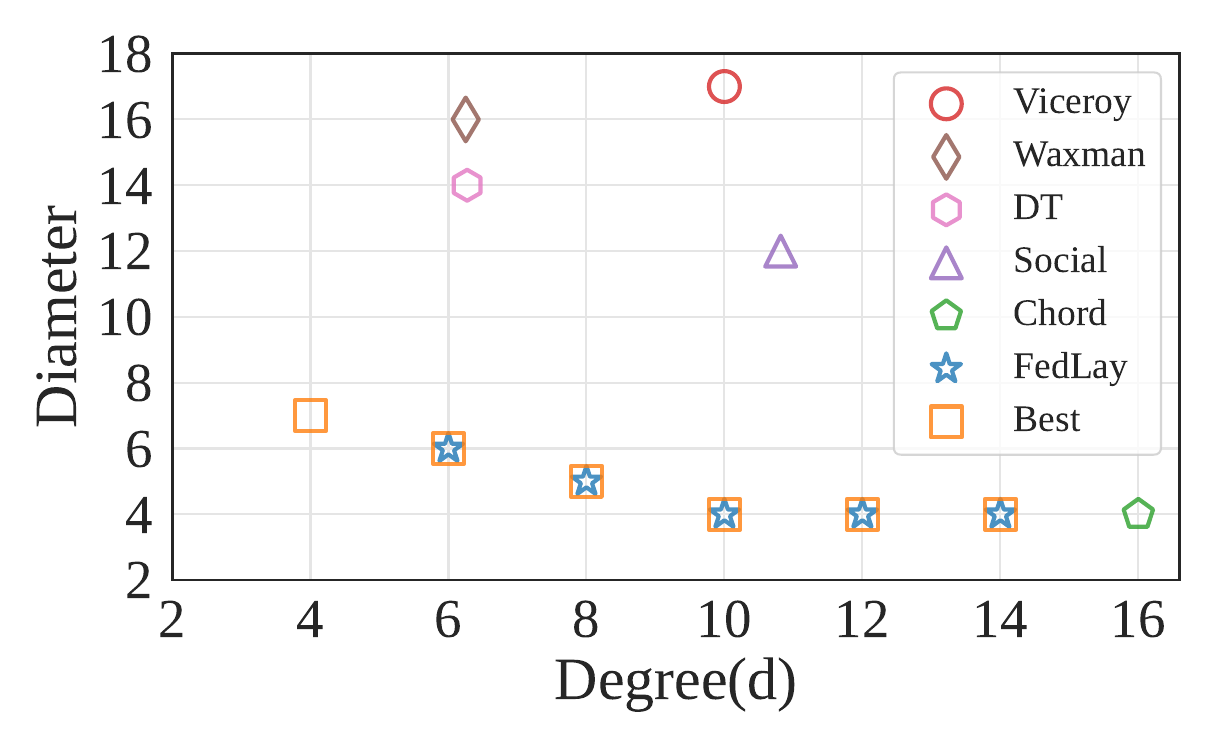}
        \vspace{-5ex}
        \caption{Network diameter}
        \label{fig:expander1-b}
    \end{subfigure}
    \begin{subfigure}[b]{0.30\textwidth}
        \centering
        \includegraphics[width=\textwidth]{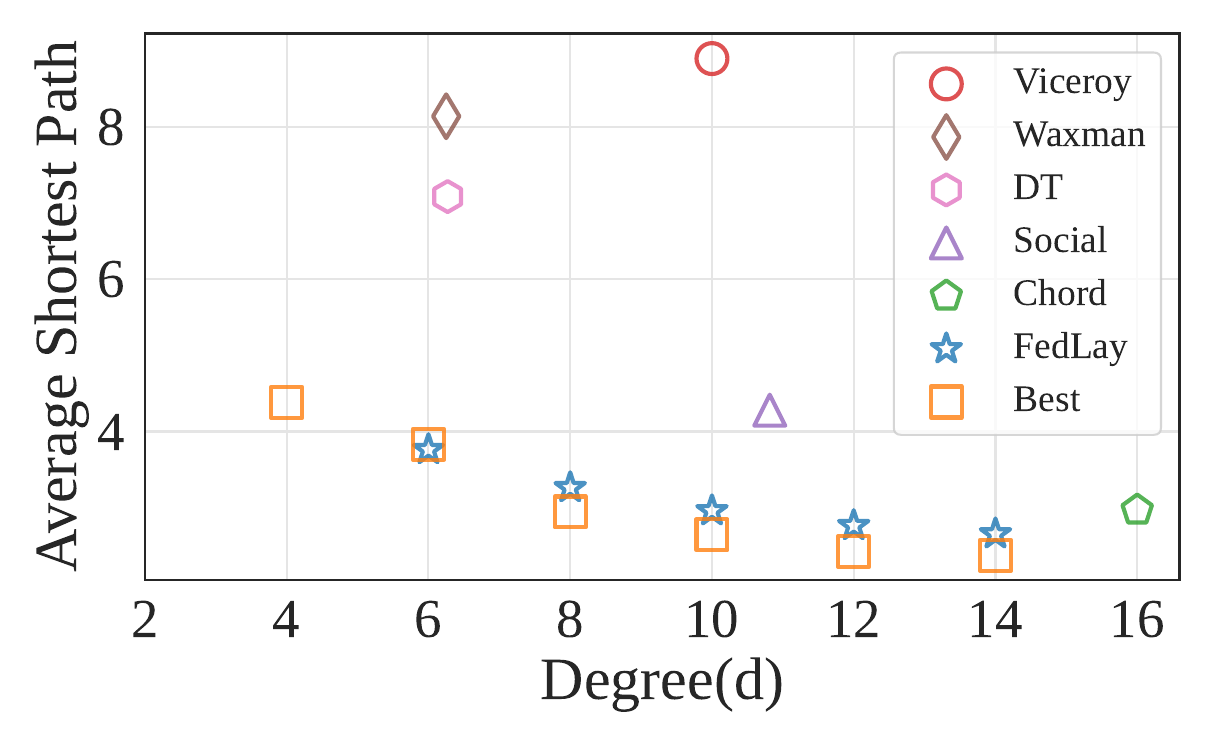}
       \vspace{-5ex}
        \caption{Average shortest path length}
        \label{fig:expander1-c}
    \end{subfigure}
   \caption{Comparisons of network topologies on the three metrics discussed in Sec.~\ref{sec:metrics}.}
   \label{fig:expander1}
    \vspace{-1ex}
\end{figure*}

The detailed decentralized construction and maintenance will be discussed in the next section. We now show that the FedLay topology is close to the optimal choice with a given node degree $d$, measured by the three metrics discussed in Section \ref{sec:metrics}. We evaluate the FedLay topology by comparing it with the following existing topologies. 

\begin{enumerate}[leftmargin=*]
\vspace{-1ex}
    \item \textbf{Best of 100 randomly generated $d$-regular graphs (``Best'').} We randomly generate 100 $d$-regular graphs and measure them for the three metrics. We obtain the optimal value among the 100 graphs for each metric. It is then considered the optimal value of practical topologies.
    for this metric. Of course, there is no decentralized protocol to construct such ``Best'' topology. 
    \item \textbf{Chord \cite{chord}.} Chord is a well-known peer-to-peer overlay network serving the function of a distributed hash table (DHT). It has a $O(\log n)$ degree and can be constructed and maintained by decentralized protocols. 
    \item \textbf{Viceroy \cite{Viceroy}.} Viceroy is a peer-to-peer overlay network with a constant degree, inspired by the classic Butterfly network used for super-computing. Its main objective is to minimize congestion by overlay routing. It can be constructed and maintained by decentralized protocols. 
    \item \textbf{Distributed Delaunay Triangulation (DT) \cite{DT-ICNP,MDT}.} DT is an overlay network with a constant degree that supports greedy routing. It can be constructed and maintained by decentralized protocols. 
    \item \textbf{Waxman network \cite{waxman}.} Waxman is a network 
    that simulates connections with 
    physical proximity.  Nodes with a close geographic distance are 
    likely to connect. There is no known decentralized construction of Waxman.  
    \item \textbf{Social network \cite{social-ego}.} We use a social network topology of Facebook users that was collected by \cite{social-ego}. This is a typical example of  overlay networks that rely on information from other application channels.
\end{enumerate} 

Fig. \ref{fig:expander1} shows the empirical results by  comparing all the above-mentioned topologies on the three metrics: the convergence factor, diameter, and average length of shortest paths. For all metrics, smaller values are more desired. Each network includes 300 nodes for fair comparisons. 
For the social network, we sample 300 nodes for fair comparisons.
We vary the node degree from 4 to 14 for both ``Best'' and FedLay. Other networks do not support flexible node degrees, so the result of each topology is shown as a single dot in each figure. 

We summarize our findings from these results as follows. 
``Best'' always provides optimal results for every metric. 
The results of FedLay are extremely close to ``Best'': most points of FedLay are superposed with those of ``Best'' with only a few exceptions. All of topologies are much less optimal compared to FedLay. The convergence factor of Chord is very high but the diameter and average shortest path length are low, due to its high node degree. 

The results of both DT and Waxman are much less optimal compared to FedLay. The main reason is that both topologies are constructed on neighbors with short distances. Hence, it might take a long latency for information (local models) to propagate between two remote nodes.  The convergence factors of both Viceroy and the social network are close to that of FedLay, but their diameter and average shortest path length are much longer than those of FedLay. The node degree of Chord is larger than most other constant-degree networks, because it needs a node degree of $2\log n$. The convergence factor of Chord is very high but the diameter and average shortest path length are low, due to its high node degree. 

Since ``Best'' are only simulated optimal values rather than realistic network topologies, 
\textbf{FedLay achieves the best results on all three metrics among existing practical overlay topologies.}

\vspace{-1ex}
\begin{definition}[A correct FedLay overlay]
    We define that a FedLay network is correct, if every node $u$ has a neighbor set $N_u$ such that $N_u$ includes the adjacent nodes of $u$ in all $L$ virtual ring spaces and does not include other nodes.  Each node also knows the virtual coordinates of all its neighbors. 
\end{definition}

FedLay uses random coordinates to achieve near-random sampling of other nodes and hence generates a near-random-regular graph.


\textbf{Understanding the FedLay topology.} Since all coordinates in the FedLay topology are randomly computed. In each virtual ring space, the two adjacent nodes of a node $u$ are actually randomly sampled from the set of all nodes. Therefore, all neighbors of $u$ in FedLay are randomly selected and all other nodes have approximately equal likelihood to be selected. 
Hence, FedLay can approximate an RRG, which, as studied in past research \cite{Jellyfish,S2-ICNP,S2-TPDS,hua2022efficient}, provides optimal results on both convergence factors and shortest paths.
So why are the coordinates in virtual spaces necessary?
The main difficulty of generating an RRG in a decentralized manner is that a node cannot find neighbors by randomly sampling all other nodes with equal likelihood if it does not know the whole network.  
The key idea of overcoming such difficulty is to use the random coordinates to order all nodes in each virtual ring space. The ring coordinates allow every newly joining node to use \textit{greedy routing} to find its two adjacent nodes in every virtual space within a small number of routing hops. Such an adjacent-node-discovery process achieves near-random sampling of other nodes. In addition, the coordinates also help the network to recover from node failures. 

\vspace{-1ex}

\section{Design of FedLay Protocols}
\label{sec:design}

\vspace{-1ex}
\subsection{Overview}
\vspace{-1ex}
\begin{figure}
    \centering
    \includegraphics[width=0.42\textwidth]{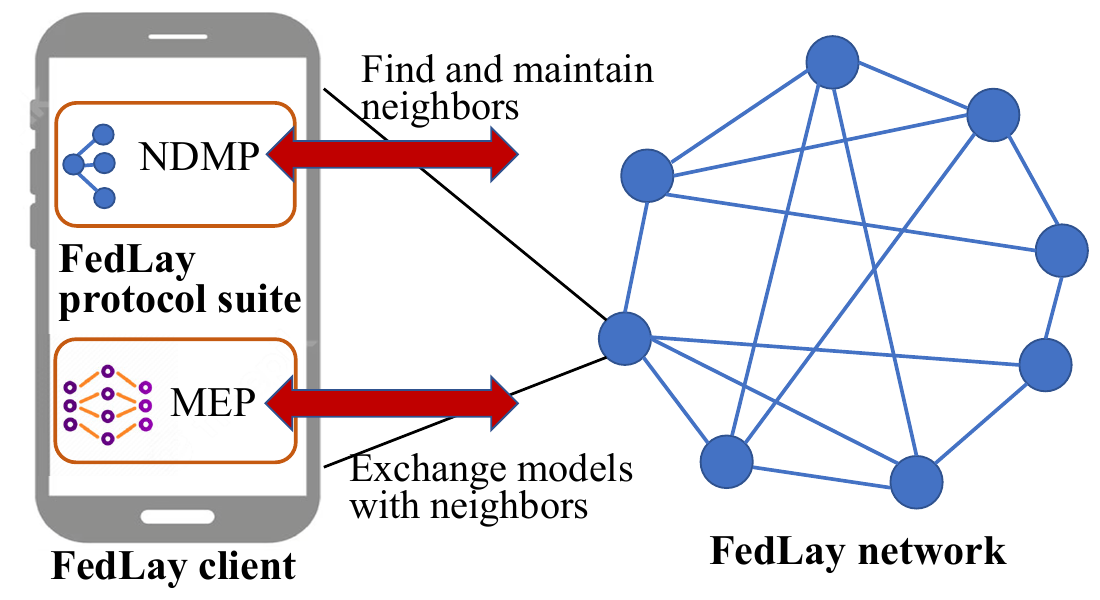}
    \caption{FedLay protocol suite includes two sets of protocols: 1) Neighbor Discovery and Maintenance Protocols; 2) Model Exchange Protocol.}
    \label{fig:protocols}
    \vspace{-3ex}
\end{figure}

\textbf{The key features of FedLay are that it is completely decentralized and all nodes are self-organized: the FedLay protocol suite allows nodes to join, leave, and fail in the network while still maintaining a correct topology and every node only stores its neighbor information. }

As shown in Fig.~\ref{fig:protocols}, the FedLay protocol suite running on a client consists of two sets: 1) Neighbor Discovery and Maintenance Protocols (NDMP); and 2) Model Exchange Protocol (MEP). 
Both sets of protocols  exchange messages with other nodes in the network.  The objective of NDMP running on node $u$ is to allow $u$ to find its correct neighbors 
during the join of $u$ and maintain correct neighbors under network dynamics. Hence, NDMP can be considered as \textit{control protocols} to construct the correct FedLay network. 
The objective of MEP is to decide when to exchange the local models with the neighbors and how to process the received models. Hence, MEP can be considered as an \textit{application protocol} to optimize the model convergence of DFL. Both NDMP and MEP use TCP with reliable delivery. 



\vspace{-1ex}
\subsection{Neighbor Discovery and Maintenance Protocols (NDMP)}

NDMP includes $\mathtt{join}$, $\mathtt{leave}$, and $\mathtt{maintenance}$ protocols. 
The $\mathtt{join}$ protocol is run by each new node joining the FedLay network. It ensures all nodes will find the correct neighbors after the node joins the network. 
The $\mathtt{leave}$ protocol is run by each node that is preparing to leave the network. It ensures all remaining nodes will keep the correct neighbors after the leave. 
The $\mathtt{maintenance}$ protocol is run by every node periodically to detect potential failed neighbors or wrong neighbors and fix these errors. 

\subsubsection{NDMP join protocol}
\label{sec:joinprotocol}

The $\mathtt{join}$ protocol is designed to achieve the following correctness property: given an existing FedLay overlay network, a new joining node runs the $\mathtt{join}$ protocol. When the $\mathtt{join}$ protocol finishes, the joining node is guaranteed to find its correct neighbors, i.e., the adjacent nodes in all virtual spaces. 
The above property will be proved later and it ensures that a correct FedLay overlay can be achieved after every new node join.  
If a FedLay network with $n$ node is correct, after a new node joins, the $(n+1)$-node network is also correct. 
This provides a way to \textit{recursively} construct a large overlay network correctly  from  a small-scale network even with 2 or 3 nodes. 
Note that such recursive property is a key module  to  ensure the correctness of many P2P overlays such as Chord \cite{chord} and distributed DT graph \cite{DT-ICNP,MDT}. 

\RE{FedLay builds upon the concept of random virtual coordinates and circular distance introduced by SpaceShuffle \cite{S2-ICNP}, extending it to achieve a fully decentralized topology construction. Unlike \cite{S2-ICNP}, where an administrator is required to access each node and switch, FedLay allows any client to join the network through any existing node, enhancing scalability and ease of use. Additionally, we optimize the NDMP leave protocol to minimize overhead by ensuring that maintenance operations are only triggered when necessary, thus reducing the resource consumption during network changes.}

When a new node $u$ joins the existing correct FedLay network, we assume $u$ knows one existing node $v$ in the overlay, which can be an arbitrary node -- this is the minimum assumption for any overlay network. If $u$ knows no existing node, it has no way to join any overlay. $u$ first computes a random coordinate as its position in the first virtual space, say $x_1^u$.
$u$ will let $v$ sends a message \textit{Neighbor\_discovery}  to the current Fedlay network using \textit{greedy routing} to the destination location $x_1^u$. \textit{Neighbor\_discovery} also includes $u$'s IP address. 

We first define the concept of \textit{circular distance}, which is a metric used in greedy routing of FedLay. 
\vspace{-1ex}
\begin{definition}[Circular distance]
    The circular distance of two coordinates $x$ and $y$ in the same ring space, $0\leq x, y < 1$, is:
\vspace{-1ex}
\[\footnotesize
CD(x,y)=\min\{|x-y|, 1-|x-y|\}.\]\
\vspace{-5ex}
\end{definition}
For two coordinates $x$ and $y$ on a ring, the circular distance is the length of the smaller arc between $x$ and $y$, normalized by the perimeter of the ring that is 1. 
We say $x$ is \textit{closer} to $y$ than $w$ on a ring space, if $CD(x,y)<CD(w,y)$. If $x$ and $w$ have the same circular distance to $y$, we always break the tie to one of $x$ and $w$ with a smaller value of their IP addresses (considering each IP address is a 32-bit value). Hence, there is only one node that is closest to a given coordinate $x$. 

Upon receiving \textit{Neighbor\_discovery}, the greedy routing protocol to the destination location $x_i^u$ in Space $i$ is executed by a  node $v$ as following: 
\begin{enumerate}[leftmargin=*]
    \item Node $v$ finds a neighbor $w$, such as $w$'s coordinate in Space $i$, $x_i^w$, has the smallest circular distance to $x_i^u$ among all neighbors of $v$. 
    \item If $CD(x_i^u, x_i^v)>CD(x_i^u, x_i^w)$, $v$ forwards the \textit{Neighbor\_discovery} message to $w$. 
    \item  If $CD(x_i^u, x_i^v)\leq CD(x_i^u, x_i^w)$, ${Neighbor\_discovery}$ stops at $v$. 
    From $v$'s two adjacent nodes, $v$ finds the adjacent node $p$ such that $x_i^u$ is on $\wideparen{v, p}$, the smaller arc between $v$ and $p$.
    Then $v$ sends a message to $u$ to tell $u$ that $v$ and $p$ are $u$'s adjacent nodes on this virtual ring and let $u$ add $v$ and $p$ to $u$'s neighbor set. 
\end{enumerate}

The greedy routing presented above will make each node forward \textit{Neighbor\_discovery} to its neighbor that has the shortest circular distance to the destination 
location $x_i^u$. When a node $v$ cannot find a neighbor that is closer to $x_i^u$ than $v$ itself, $v$ must be the node that has the shortest circular distance to $x_i^u$ among all nodes in FedLay (will be formally proved later). Hence, $v$ and one of its adjacent nodes $p$ will be $u$'s neighbors.

 \begin{figure}[t]
    \centering
    \includegraphics[width=0.48\textwidth]{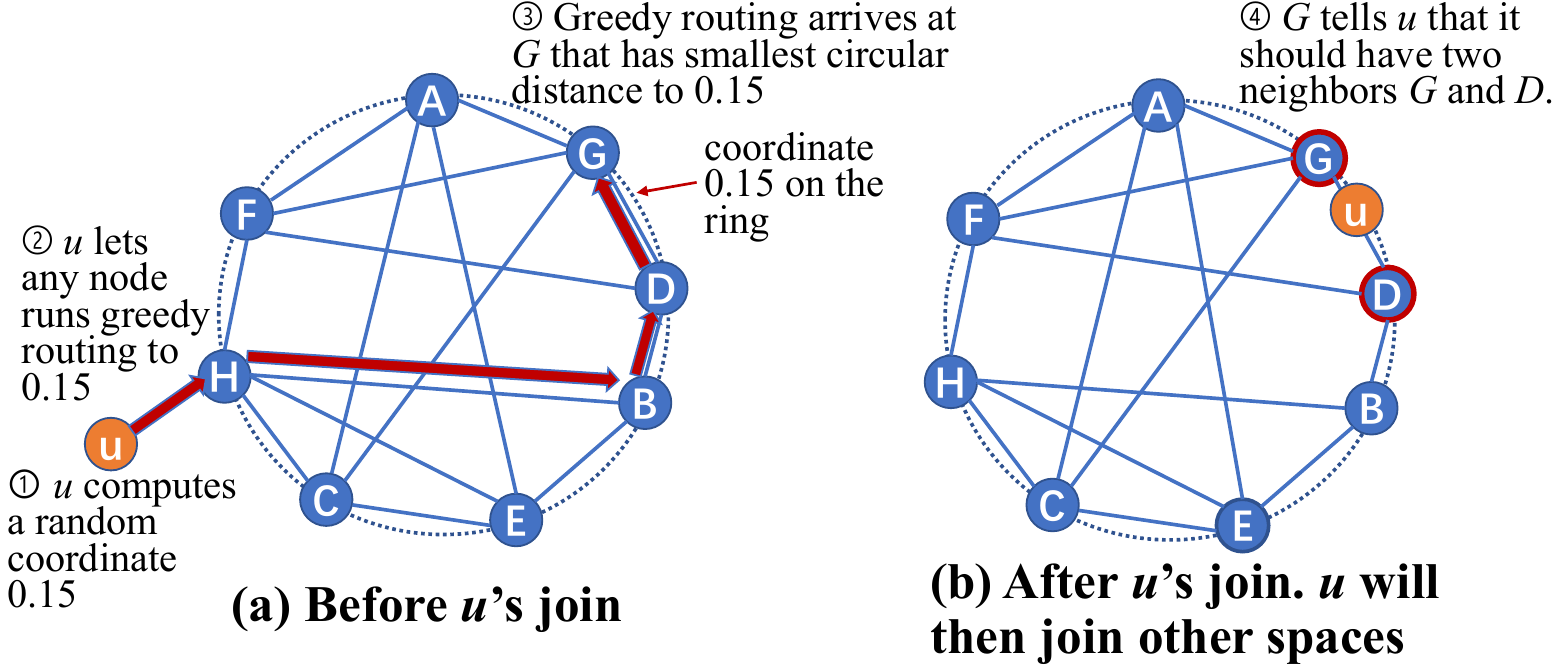}
    \vspace{-1ex}
    \caption{An example of the FedLay $\mathtt{join}$ protocol.}
    \label{fig:join-example}
    \vspace{-3ex}
\end{figure}

We show an example of FedLay $\mathtt{join}$ in Fig.~\ref{fig:join-example}. $u$ joins FedLay and it knows an existing node $H$ in the network. $u$ computes a random coordinate 0.15 and asks $H$ to run greedy routing of \textit{Neighbor\_discovery} to 0.15. $H$ will forward \textit{Neighbor\_discovery} to $B$, which is closest to 0.15 in the space among all $H$'s neighbors. Eventually, the message arrives at $G$, the node that is closest to 0.15 in the space and $G$ tells $u$ to add $G$ and $D$ as neighbors. Note greedy routing has much smaller hop-count than traveling nodes one after another through the ring because there are many shortcuts like the link $HB$.

We have the following property. 
\vspace{-1ex}
\begin{lemma}
In a ring space of a correct FedLay network and a given coordinate $x$, if a node $v$ is not the node that has the smallest circular distance to $x$ in the space, then $v$
must have an adjacent node $w$ on the ring such that $CD(x, x^v)>CD(x, x^{w})$, where $x^v$ is $v$'s coordinate.
\end{lemma}
\vspace{-1ex}
 \begin{proof}
Let $p$ be the node with the smallest circular distance to  $x$ in the space. Consider the two arcs between $x$ and $x^v$. 
At least one of them has a length no longer than 0.5. Let that arc be $\wideparen{x^v, x}$ with length $L(\wideparen{x^v, x})$. $CD(x^v, x)=L(\wideparen{x^v, x})\leq 0.5$.

If $p$ is on $\wideparen{x^v, x}$, consider the arc between $v$ and $p$, $\wideparen{x^v, x^p}$, as a part of $\wideparen{x^v, x}$.  If $v$ has an adjacent node $q$ whose coordinate 
is on $\wideparen{x^v, x^p}$, then $L(\wideparen{x, x^q})<L(\wideparen{x^v, x})\leq 0.5$. Hence, $CD(x^v,x)>CD(x^q,x)$. The lemma holds in this situation.
If $v$ has no adjacent node on $\wideparen{x^v, x^p}$, then there must be no node on the arc $\wideparen{x^v, x^p}$. Hence, $v$ and $p$ are adjacent nodes. 
Since $p$ is the node closest to $x$,  $v$ does have an adjacent node $p$ that is closer to $x$. The lemma also holds in this situation.

If $p$ is not on $\wideparen{x^v, x}$, consider the arc $\wideparen{x^v, x, x^p}$. The arc $\wideparen{x, x^p}$ is a part of $\wideparen{x^v, x, x^p}$, and we have $L(\wideparen{x, x^p})<L(\wideparen{x^v, x})$, because $p$ is the closest node to $x$. If $v$ has an adjacent node $q$ whose coordinate 
is on $\wideparen{x^v, x, x^p}$, then $L(\wideparen{x, x^q})<L(\wideparen{x^v, x})\leq 0.5$. Hence, $CD(x^v,x)>CD(x^q,x)$. The lemma holds in this situation.
If $v$ has no adjacent node on $\wideparen{x^v, x, x^p}$, then $v$ and $p$ are adjacent nodes. Since $p$ is the node closest to $x$,  $v$ does have an adjacent node $p$ that is closer to $x$. The lemma also holds in this situation.

Therefore in all cases, the lemma holds.
\end{proof}

Note that every adjacent node of $v$ is its neighbor in a FedLay network. 
This lemma tells that if $v$ is not the node that is closest to the destination coordinate $x_i^u$, then the greedy routing algorithm must execute Step 2 and forwards the message to a neighbor. 
Hence, if the greedy routing algorithm goes to Step 3 and stops at $v$, $v$ must be the node that has the smallest circular distance to $x_i^u$ in the space.
So we have, 
\vspace{-1ex}
\begin{theorem}
In a FedLay network, 
when \textit{Neighbor\_discovery} to the destination coordinate $x$ stops at a node $v$, $v$ must be the node that has the smallest circular distance to $x$.
\end{theorem}
\vspace{-1ex}
The \textit{Neighbor\_discovery} message will stop at the  node $v$ closest to $x_i^u$ and $v$ must be an adjacent node to the joining node $u$, because no other node is closer to $u$'s coordinate $x_i^u$. $v$ also knows the other adjacent node $w$ of $u$, because $w$ is a current adjacent node of $v$. 
$v$ will send a message to $u$ by TCP including the information of $v$ and $w$. $u$ will then add $v$ and $w$ to its neighbor set.

The joining node $u$ can find all its neighbors by running the above $\mathtt{join}$ protocol in all spaces. 
Therefore, \textbf{if $u$ joins a correct FedLay network, the new FedLay network after this join is also correct. }

In some extreme cases, there could be multiple nodes joining the network simultaneously. This situation will be handled by both the $\mathtt{join}$ and $\mathtt{maintenance}$ protocols.

\subsubsection{NDMP leave protocol}
 \begin{figure}[t]
    \centering
    \includegraphics[width=0.42\textwidth]{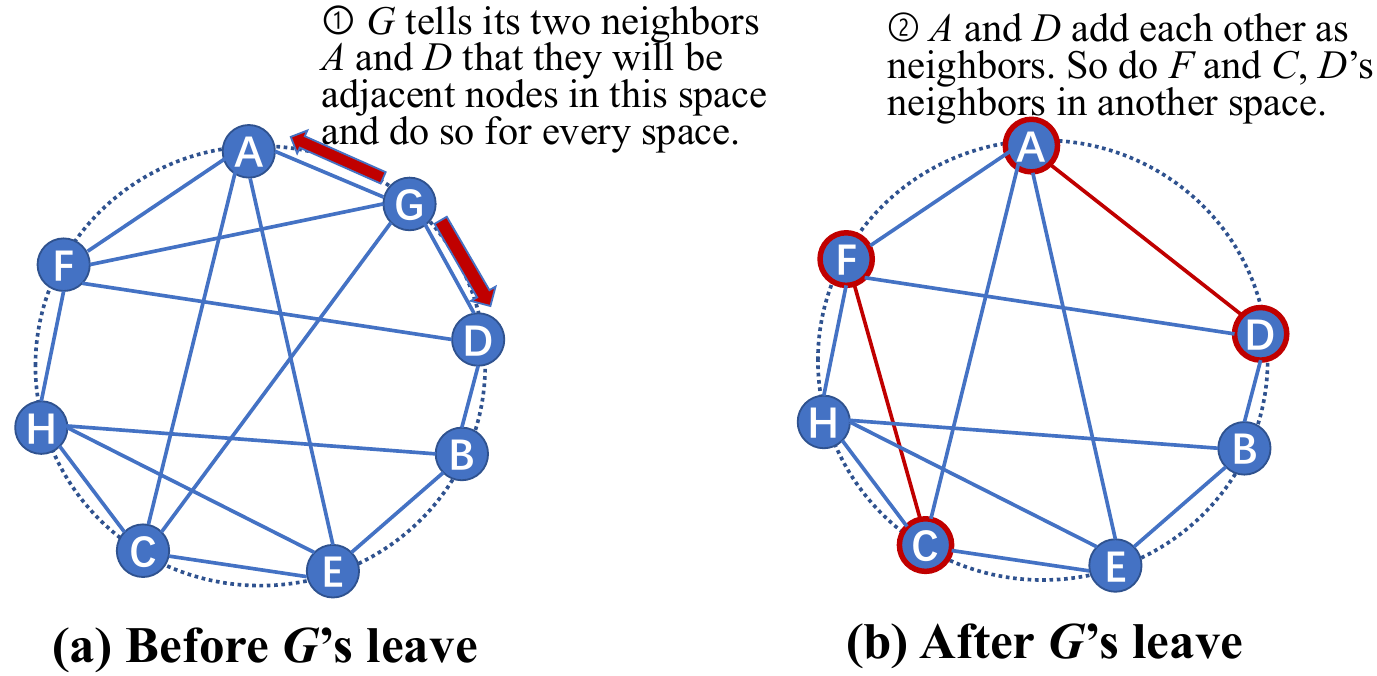}
    \vspace{-1ex}
    \caption{An example of the FedLay $\mathtt{leave}$ protocol.}
    \label{fig:leave-example}
    \vspace{-1ex}
\end{figure}

The $\mathtt{leave}$ protocol of NDMP is quite straightforward. When a user wants to leave and closes the client program of FedLay, 
for every virtual space, the leaving node sends messages to its two adjacent nodes and tells them to add each other to their neighbor sets. 
As shown in Fig.~\ref{fig:leave-example}, node $G$ wants to leave the network and tells its two adjacent nodes $A$ and $D$ about its leaving. 
Then $A$ and $D$ will consider each other as adjacent nodes and neighbors. In another space, $G$'s two adjacent nodes $F$ and $C$ will also add each other to their neighbor sets. 
Hence, \textbf{the new FedLay network after a node leave is also correct. }

 \begin{figure}[t]
    \centering
    \includegraphics[width=0.45\textwidth]{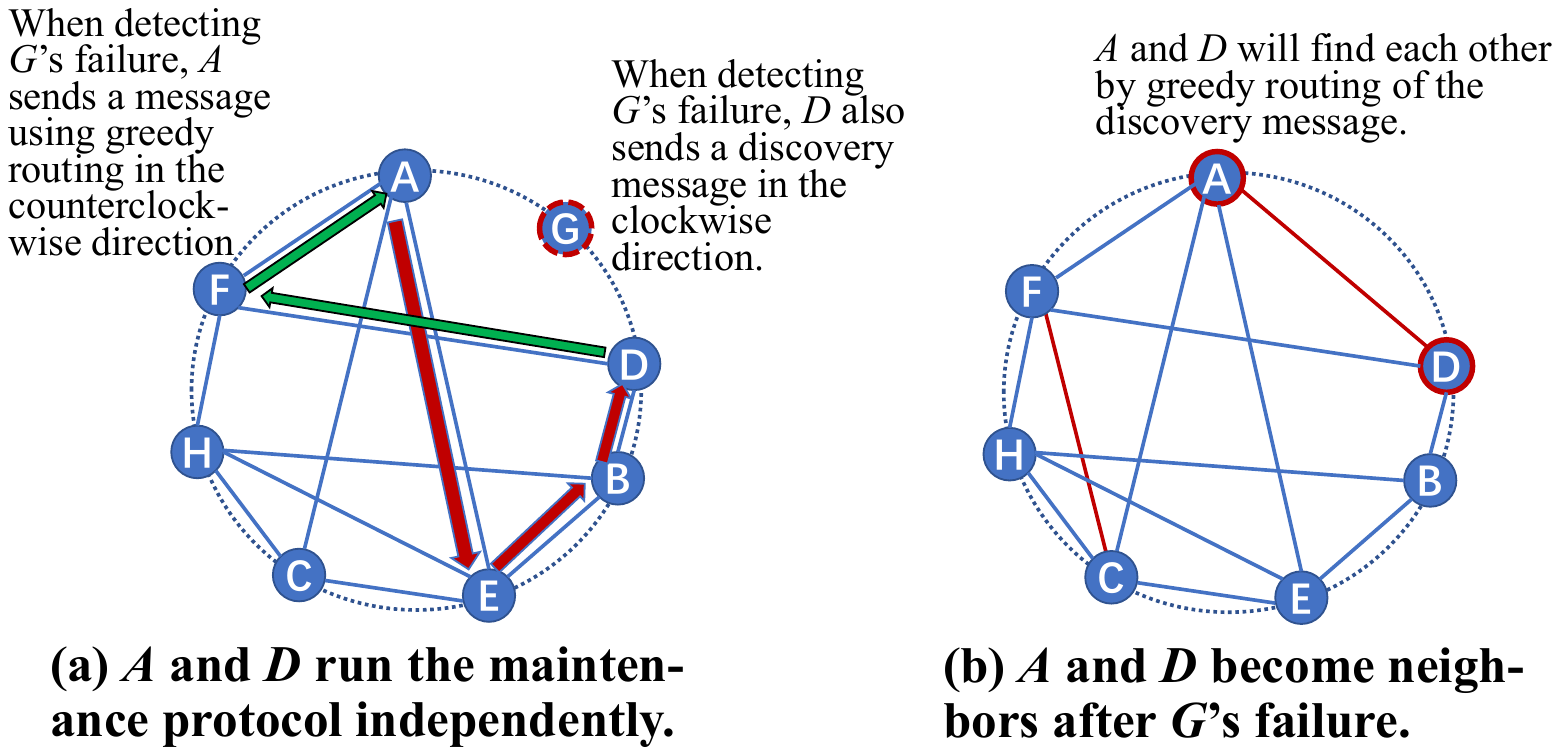}
    
    \vspace{-1ex}
    \caption{An example of the $\mathtt{maintenance}$ protocol.}
    \label{fig:maintain-example}
    \vspace{-3ex}
\end{figure}

\subsubsection{NDMP maintenance protocol}
In addition to planned  leaves, An overlay network may also 
experience node failures due to various reasons such as an Internet service outage and end system failures. 
A failed node disappears without notice. In order to detect and fix these situations, each node in FedLay also runs the $\mathtt{maintenance}$ protocol.

The $\mathtt{maintenance}$ protocol requires every node to send each of its neighbors a heartbeat message periodically. 
Suppose the time period between two heartbeat messages is $T$. If a node $p$ has not received any heartbeat message from a neighbor $u$ for $3T$ time, it considers that $u$ has failed.  
$p$ then sends a \textit{Neighbor\_repair} message by greedy routing in the opposite direction of $u$ on the virtual space $i$ where $u$ and $p$ are adjacent nodes.
We use an example to explain this protocol, as shown in Fig.~\ref{fig:maintain-example}. 
In the example of Fig.~\ref{fig:maintain-example}(a), node $A$ detects the failure of $G$. Since $G$ is an adjacent node of $A$ on $A$'s clockwise side, $A$ sends a \textit{Neighbor\_repair} message by greedy routing in the counterclockwise direction (the opposite direction of $G$). By ``counterclockwise direction'', it requires that all hops of such greedy routing, $A-E-B-D$ in this example,  should follow the counterclockwise order.

We give a  formal description of the counterclockwise direction:
Upon receiving \textit{Neighbor\_repair} to $x_i^u$ in Space $i$, a  node $v$ runs the following algorithm: 
\begin{enumerate}[leftmargin=*]

    \vspace{-1ex}
    \item Node $v$ considers a subset of its neighbors, such that for every neighbor $w$ in the subset, $w$'s coordinate in Space $i$, $x_i^w$, satisfies $x_i^w<x_i^v$ or $x_i^w>x_i^u$. 
    For each neighbor $w$, consider the length of the arc from $x_i^w$ to $x_i^u$ in the counterclockwise direction $L(\wideparen{x_i^w, x_i^u})$. 
    From the above subset,   node $v$ finds a neighbor $w'$, such that $w'$'s coordinate in Space $i$, $x_i^{w'}$, has the smallest arc length $L(\wideparen{x_i^{w'}, x_i^u})$ among all neighbors in the subset. 

    \item If $L(\wideparen{x_i^v, x_i^u})>L(\wideparen{x_i^{w'}, x_i^u})$, $v$ forwards \textit{Neighbor\_repair} to $w$. 

    \item  If $L(\wideparen{x_i^v, x_i^u})<L(\wideparen{x_i^{w'}, x_i^u})$, \textit{Neighbor\_repair} stops at $v$. 
    $v$ then  tells $p$ that  $v$ is $p$'s adjacent nodes on this virtual ring and let $p$ add $v$ to $p$'s neighbor set. 
    
\end{enumerate}

When  \textit{Neighbor\_repair} stops, it arrives at another adjacent node of $G$ before $G$'s failure (stated in a later theorem). Then $G$'s previous two adjacent nodes can be connected.  

In Fig.~\ref{fig:maintain-example}(a), node $D$ also detects $G$'s failure, independent of $A$'s detection. Then $D$ sends a \textit{Neighbor\_repair} message to the clockwise direction and the message will travel on a path $D-F-A$. 
The algorithm to forward \textit{Neighbor\_repair} in the \textit{clockwise direction} can be specified in a similar way.

By changing the subset selection condition in Step 1 to ``$x_i^w>x_i^v$ or $x_i^w<x_i^u$''. We can prove the following theorem:
\vspace{-1ex}
\begin{theorem}
Consider a correct FedLay network and a node $u$ fails. 
When a node $p$ detects the failure of its adjacent node $u$ in Space $i$, it sends \textit{Neighbor\_repair} to the destination coordinate $x_i^u$ in the opposite direction of $u$. 
When \textit{Neighbor\_repair} stops at a node $q$, $q$ is another adjacent node of $u$ in Space $i$ before $u$'s failure. 
\end{theorem}
\vspace{-1ex}

\begin{proof}
We first discuss the case that \textit{Neighbor\_repair} is sent in the counterclockwise direction. The case of clockwise direction can be proved in the same way. 

Let $q$ be another adjacent node of $u$ in Space $i$ before $u$'s failure. 

Let $v$ be the current node that receives \textit{Neighbor\_repair}. 
If $v \neq q$, $v$ has at least one neighbor $w$, such that $L(\wideparen{x_i^w, x_i^u})< L(\wideparen{x_i^v, x_i^u})$. 
Hence, \textit{Neighbor\_repair} will not stop if $v \neq q$.
$v$ will forward \textit{Neighbor\_repair} to the neighbor $w'$ and $L(\wideparen{x_i^{w'}, x_i^u})\leq L(\wideparen{x_i^w, x_i^u}) < L(\wideparen{x_i^v, x_i^u})$.

Therefore after each hop, the arc length $L(\wideparen{x_i^v, x_i^u})$ strictly decreases.
Hence, \textit{Neighbor\_repair} will not experience a loop. 
Since there are a limited number of nodes, \textit{Neighbor\_repair} will stop at $q$. 
\end{proof}

Based on the theorem, in every virtual space, the two adjacent nodes of the failed $u$ can find and  connect each other. 
\textbf{Hence, if a node fails in a correct FedLay network, after the node failure FedLay is still correct.} 

\textbf{Neighbor repair for concurrent joins and failures.} 
Note the above property cannot be proved for multiple failures that happen at the same time, called concurrent failures. 
For concurrent joins and failures, we allow each node $u$  to periodically send two  \textit{Neighbor\_repair} messages with destination $x_i^u$ to both counterclockwise and clockwise directions in every virtual space $i$,
even without detecting any neighbor failure.
For each node $v$ in  $u$'s neighbor set, if they  are indeed adjacent in a virtual space, $v$ will receive a \textit{Neighbor\_repair} message that stops at $v$. 
The \textit{Neighbor\_repair} stops at a node $w$ that is not in $u$'s neighbor set, $w$ and $u$ will add each other to their neighbor sets. 
In fact there is no way to prove any property for concurrent joins and failures in any structured P2P network \cite{chord,MDT}.
Hence we conduct experiments of extreme concurrent joins and failures and the above method always allows the network to recover to a correct FedLay. 

\vspace{-1ex}
\subsection{Model Exchange Protocol (MEP)} 
\vspace{-.5ex}

One key challenge of DFL is that there is no central server to evaluate the quality of models from different clients. 
In P2P model exchanges, a client with low-quality local models can `infect' its neighbors with high-quality models and these errors may be further propagated in the overlay. 
MEP is designed to limit the impact of low-quality models and amplify the impact of high-quality models in a decentralized way. 
We consider two practical issues in DFL systems. 1) Data heterogeneity \cite{he2018cola,Beltran2022DFL,hua2022efficient}. It is well known that the local data of different clients are usually non-iid due to geographic and environmental diversity. Hence, their models have different accuracy. 
2) Client heterogeneity \cite{zehtabi2022decentralized,cao2021hadfl, pang2022incentive}. Clients of DFL could 
have different bandwidth and computing capacities. 
They may have different model exchange frequencies. 
Unlike previous work of DFL that usually assumes homogeneous clients \cite{he2018cola,elgabli2020communication,bellet2022d,vogels2022beyond,chow2016expander,hua2022efficient}, 
 MEP allows each client to set different parameters based on their two dimensions of heterogeneity. to guide clients to exchange models with their neighbors, including multiple design components. We present three main components in detail:  1) set confidence parameters; 2) set model exchange period;  3) model de-duplication.

\begin{table*}[t]
\fontsize{9}{9}\selectfont
    \centering
    \renewcommand\arraystretch{1.12} 
    \begin{tabular}{|c|c|c|c|c|}
        \hline
        \textbf{Dataset} & \textbf{Tasks} & \textbf{Model} & \textbf{Model size per client} & \begin{tabular}{@{}c@{}}
            Comm. period for\\[-1ex] medium-cap. clients  \end{tabular} \\
        \hline
        MNIST & Img Classification & MLP & 247 KB & 5 min \\
        \hline
        CIFAR-10 & Img Classification & CNN & 1.1 MB & 10 min \\
        \hline
        Shakespeare & Next-character pred. & LSTM & 23.4 MB & 40 min \\
        \hline
    \end{tabular}
    \caption{Datasets used in evaluation}
    \label{tab:expset}
\end{table*}

\vspace{-.5ex}
\subsubsection{Asynchronous model exchange}

Previous work assumes synchronous, round-based communication, in which all clients use the same time period to exchange models with neighbors \cite{he2018cola,elgabli2020communication,bellet2022d,vogels2022beyond,chow2016expander,hua2022efficient}. However, 
due to client heterogeneity, some low-resource clients may become ``stragglers'' that will fail to perform model exchanges in the given time period, while powerful or newly joined clients prefer 
shorter time period (or higher frequency) of model exchanges.

MEP uses asynchronous communication and allows each client $u$ to use a different communication time period $T_u$. 
$T_u$ can be set in two ways: \textbf{1) Coarse-grained settings.} Each client may configure a period based on their device and communication types,
for example, $\mathtt{Server}$-$\mathtt{LAN}$, $\mathtt{PC}$-$\mathtt{LAN}$, $\mathtt{Laptop}$-$\mathtt{WLAN}$, $\mathtt{Phone}$-$\mathtt{LTE}$, and $\mathtt{IoT}$-$\mathtt{WLAN}$. 
These values are pre-specified in the client program. \textbf{2) Fine-grained settings.} Based on the monitoring of available bandwidth and computing resources,  client $u$ estimates the minimum time duration $T_{u,{\min}}$ to produce an updated ML model and transmit it to all neighbors. Then its communication period $T_u=\eta T_{u,{\min}}$ for constant $\eta>1$.

For two neighbors with periods $T_u$ and $T_v$, their model exchange period is set to $\max(T_u, T_v)$. Hence, a client may have different exchange periods to different neighbors. 

\begin{figure*}
    \centering
    \begin{subfigure}[b]{0.30\textwidth}
        \centering
        \includegraphics[width=\textwidth]{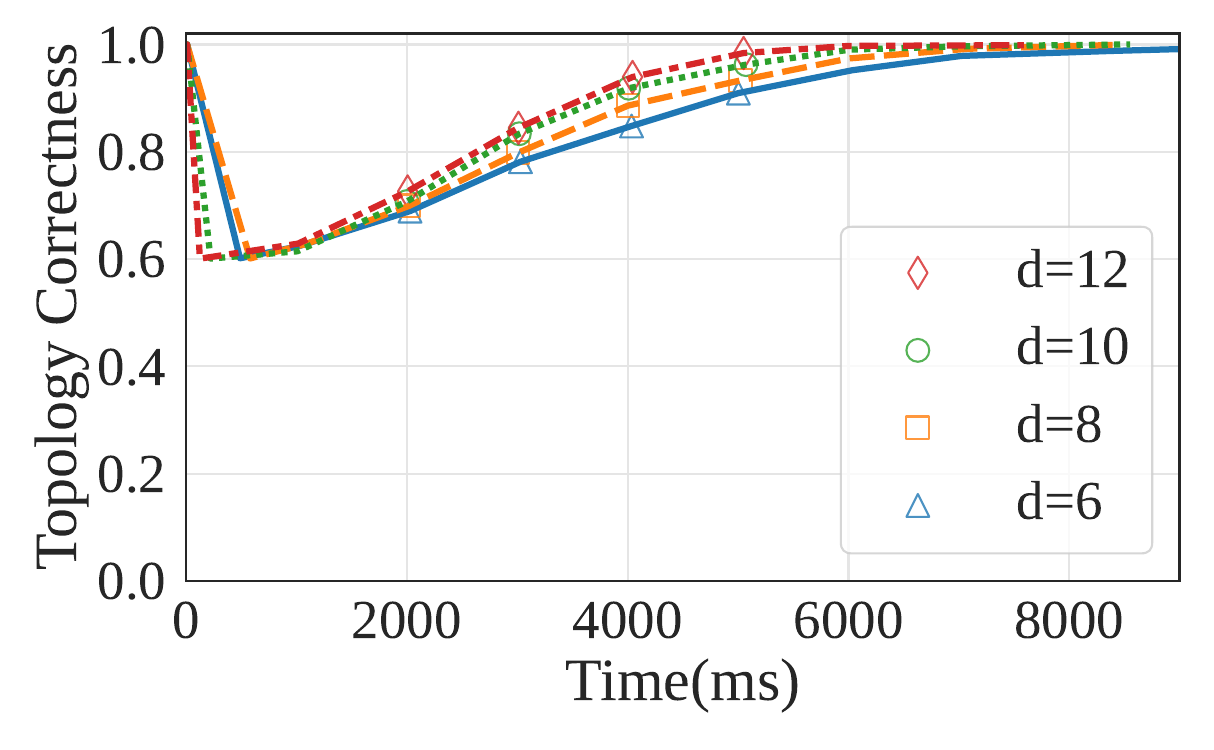}
        \vspace{-4.2ex}
        \caption{\footnotesize Correctness for concurrent joins}
        \label{fig:expander4}
    \end{subfigure}
    \begin{subfigure}[b]{0.30\textwidth}
        \centering
        \includegraphics[width=\textwidth]{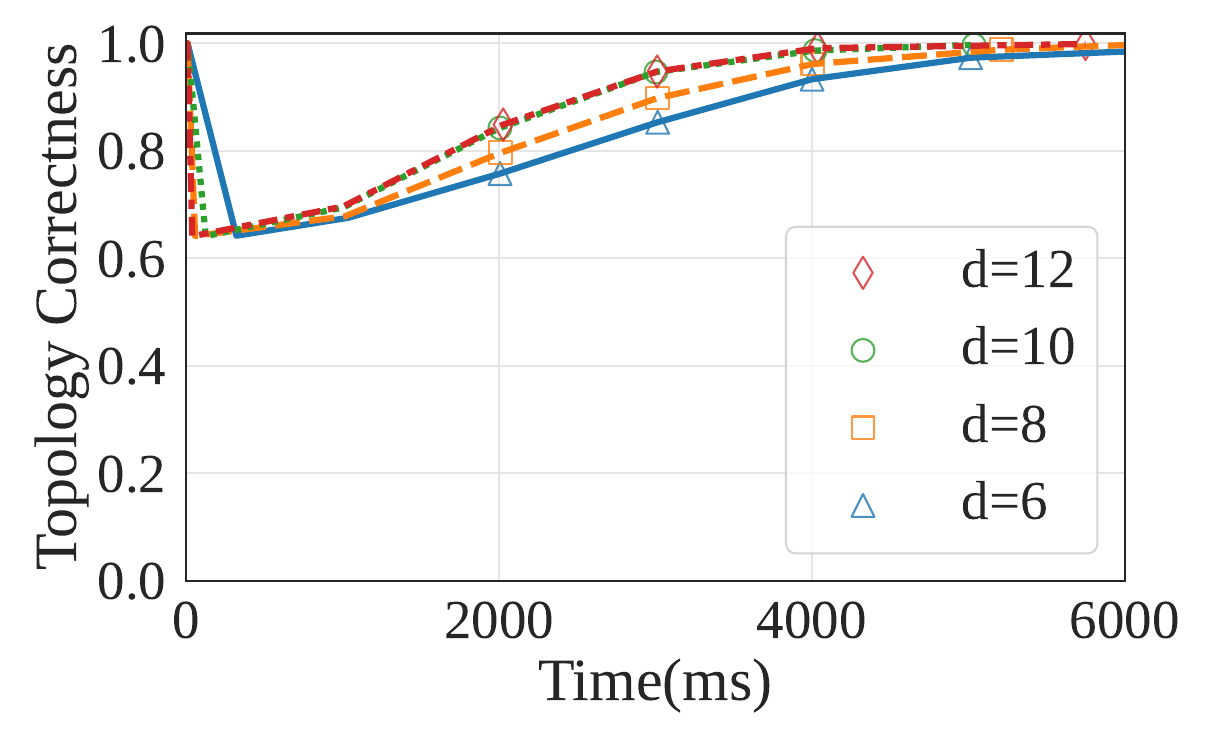}
        \vspace{-4.2ex}
        \caption{\footnotesize Correctness for concurrent fails}
        \label{fig:expander5}
    \end{subfigure}
    \begin{subfigure}[b]{0.30\textwidth}
        \centering
        \includegraphics[width=\textwidth]{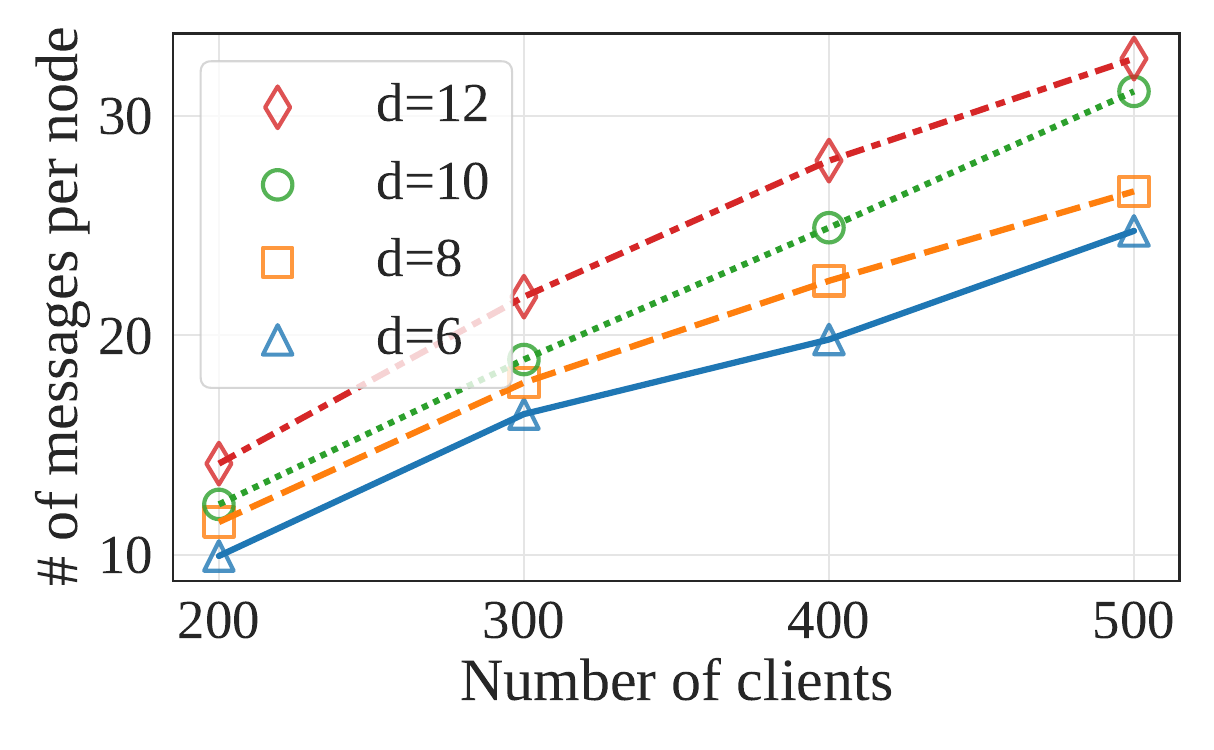}
       \vspace{-4.2ex}
        \caption{\footnotesize Message cost}
        \label{fig:expander6}
    \end{subfigure}
    \vspace{-1ex}
   \caption{Topology correctness under churn \& message cost}
   \vspace{-2ex}
   \label{fig:expander-dynamic}
\end{figure*}

\subsubsection{Set confidence parameters}

One key innovation in MEP is to introduce \textit{confidence parameters}. Each node has a set of confidence parameters that present its self-evaluation of the local model accuracy. 

We 
define the data divergence confidence $c_d$ on a client $u$:
\vspace{-2ex}
\[\footnotesize
c_{d}^u = \frac{1}{\exp(DKL(D_{loc} || D_{std}))}\]
where $DKL(\cdot)$ is the Kullback-Leibler divergence \cite{Kullback1951divergence}  to evaluate the statistical distance between two probability distributions $P$ and $Q$, $D_{loc}$ denotes the local data distribution, and 
$D_{std}$ denotes the estimated iid distribution of the dataset. The uniform distribution is widely used \cite{flniid, sattler2019robust} to estimate the iid data because the majority of publicly-available datasets for classification follow uniform distributions, such as (MNIST \cite{lecun2010mnist} and CIFAR-10 \cite{cifar_09}).
The Kullback-Leibler divergence can effectively represent the richness of a local dataset. $c_{d}\in (0,1]$ and a higher value represents a higher quality of local data and local models. 

In addition, we define the communication confidence $c_c$ on a client $u$:
$c_{c}^u = \frac{1}{T_u}$.
The intuition of using $c_c$ is that when a client has more frequent model exchanges with its neighbors, its models are more likely to have higher qualities. 

Hence, the overall confidence of client $u$ is
\vspace{-1ex}
\[  \footnotesize
c^{u} = \alpha_{d}\frac{c_{d}^{u}}{\max(c_{d})} + \alpha_{c}\frac{c_{c}^{u}}{\max(c_{c})} \]
where $\max(c_{d})$ and $\max(c_{c})$ are the maximum values of $c_{d}$ and $c_{c}$ respectively, from all $u$'s neighbors. 
$\alpha_{d}$ and $\alpha_{c}$ are two constants to balance the weights of the two confidence parameters. 
The specific values of $\alpha_{d}$ and $\alpha_{c}$ can just be 0.5 and 0.5. We try a variety of combinations of $\alpha_{d}$ and $\alpha_{c}$, and in all cases, FedLay achieves fast model convergence on different nodes. 

The models from $u$'s neighbors are aggregated  as follows:
\vspace{-1ex}
\[\footnotesize
\omega^u =  \frac{\sum_{j\in N\bigcup\{u\}} c^j\omega^j}{\sum_{j\in N\bigcup\{u\}}c^j}\]

The above aggregation will be computed once every local time period $T_u$ and the models from each neighbor are always the most updated ones from the neighbor.
In this way, clients with low confidence in their model accuracy will have less impact on other clients. In this work, we do not consider the situation where a client might intentionally set a large confidence value to mislead other clients.  

\subsubsection{Model fingerprinting}
In the neighbor set of each client, other than the IP addresses, coordinates, and confidence values of the neighbors, the client also stores the fingerprint $f$ of the most recent models of each neighbor, computed by hashing the models by a public hash function.  
Before starting a model exchange, the client first sends the fingerprint to its neighbor. If the neighbor finds the fingerprint matches the models that are going to be sent, the neighbor will consider the models to be a duplicate of a copy sent earlier and stop sending the models. This method reduces unnecessary traffic of exchanges of duplicated  models.  

\section{Performance Evaluation}
\label{sec:evaluation}

\subsection{Evaluation methodology}

\subsubsection{Three types of evaluation}
We conduct three types of evaluation of FedLay for different scales of DFL networks. 


\textbf{1) Real experiments.} We conduct experiments with real packet exchanges and data training. 
We deploy 16 instances to public clouds (we used both Oracle OCI and Amazon EC2), each with a 2GHz CPU and 2GB RAM. Each instance is connected to the Internet  and runs a FedLay client. 
Each client sends and receives NDMP and MEP messages using TCP. 
Clients train ML models on their local datasets with Pytorch \cite{pytorch} and exchange the models with active neighbors. 
There is no central server for any purpose, and the system is completely decentralized. 
The purpose of this type of experiment is to present a prototype and demonstrate that FedLay can run with real ML data training in practice. 

\textbf{2) Medium-scale emulation with real data training.} In this type of experiment, we use real data training and simulated packet exchanges as discrete events to evaluate networks with up to 100 clients. 
The simulation and real data training and testing run on a machine with an NVIDIA GeForce RTX3080 graphic card for training acceleration. 
The purpose of this type of experiment is to evaluate the overlay construction/maintenance, model accuracy, convergence speed, and message cost of FedLay and other DFL methods. 

\textbf{3) Large-scale simulation with trained models.} 
For more than 100 clients, conducting all data training on a few machines takes very long time. For networks with 200 to 1000 nodes, we re-use the models trained from the above two types of experiments and assign them to the simulated clients. Packet exchanges are simulated as discrete events. 

\begin{figure*}
    \centering
    \begin{subfigure}[b]{0.30\textwidth}
        \centering
        \includegraphics[width=\textwidth]{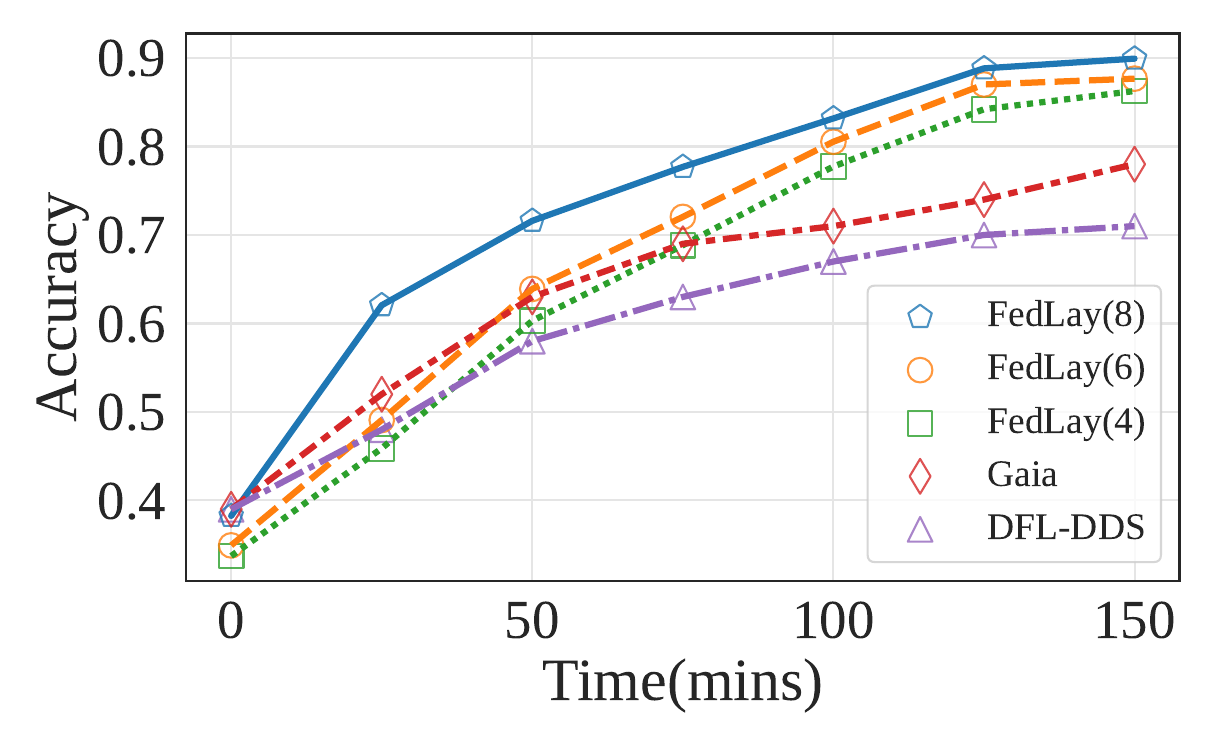}
        \vspace{-4.2ex}
        \caption{\footnotesize MNIST. Accuracy vs time (mins)}
        \label{fig:acc-m-n16s3}
    \end{subfigure}
    \begin{subfigure}[b]{0.30\textwidth}
        \centering
        \includegraphics[width=\textwidth]{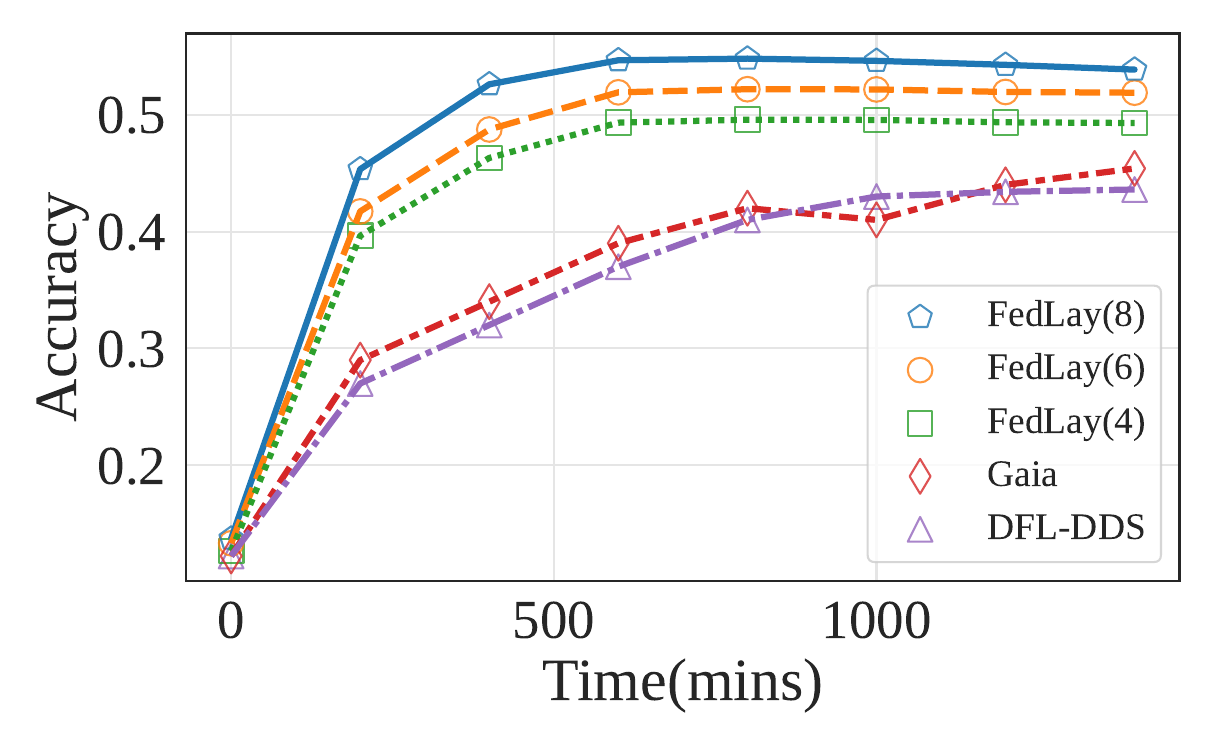}
        \vspace{-4.2ex}
        \caption{\footnotesize CIFAR-10. Accuracy vs time (mins)}
        \label{fig:acc-c-n16s8}
    \end{subfigure}
    \begin{subfigure}[b]{0.30\textwidth}
        \centering
        \includegraphics[width=\textwidth]{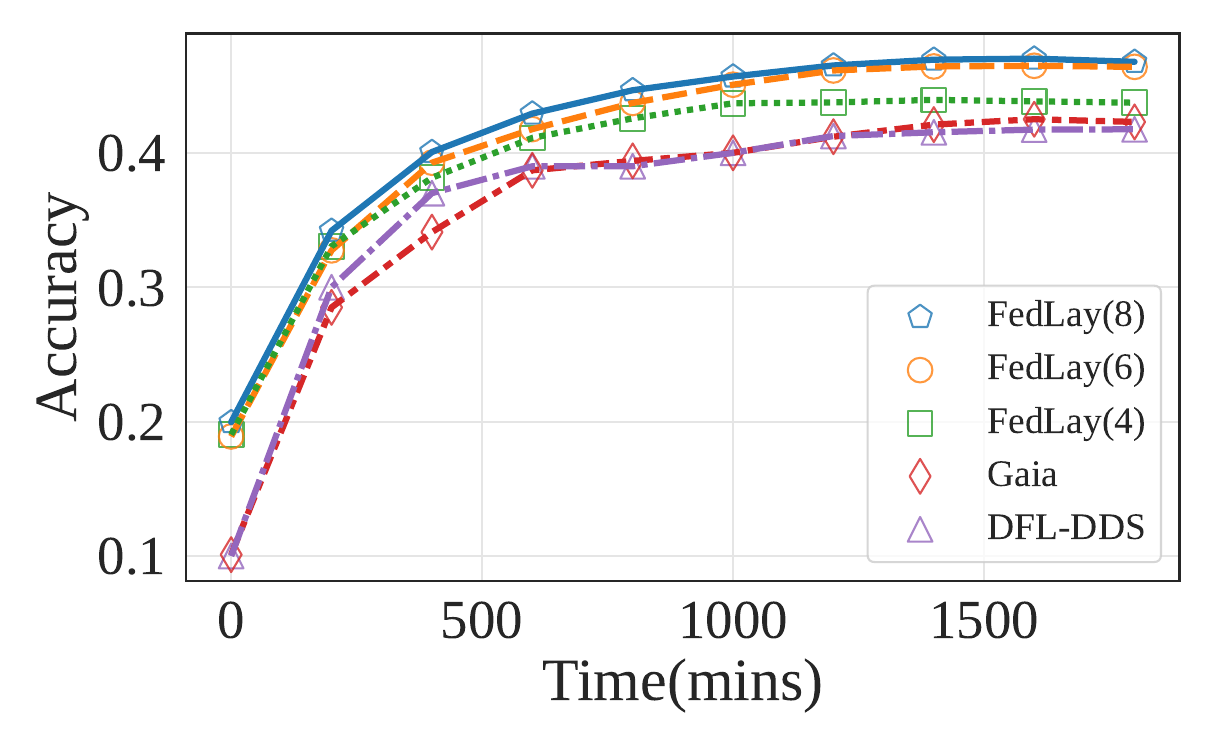}
        \vspace{-4.2ex}
        \caption{\footnotesize Shakespeare. Accuracy vs time (mins)}
        \label{fig:acc-s-n16}
    \end{subfigure}
    \begin{subfigure}[b]{0.30\textwidth}
        \centering
        \includegraphics[width=\textwidth]{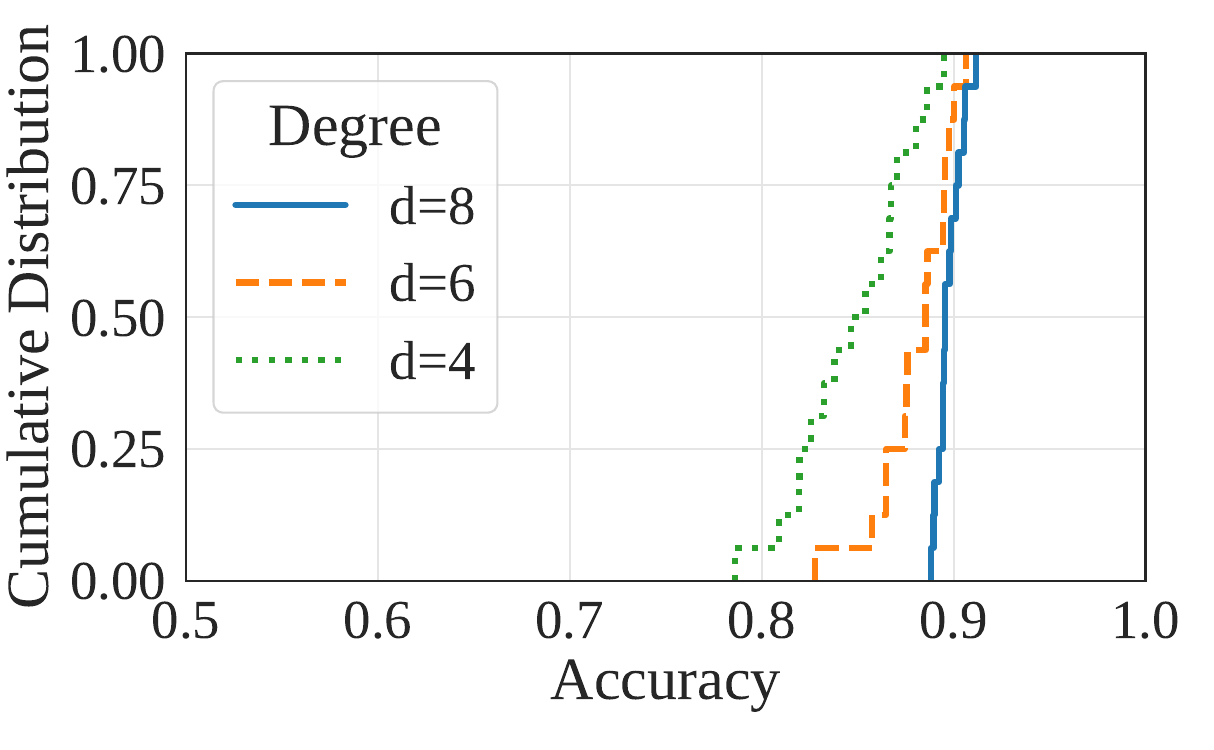}
        \vspace{-4.2ex}
        \caption{\footnotesize MNIST. CDF of Accuracy}
        \label{fig:cdf-m-n16s3}
    \end{subfigure}
    \begin{subfigure}[b]{0.30\textwidth}
        \centering
        \includegraphics[width=\textwidth]{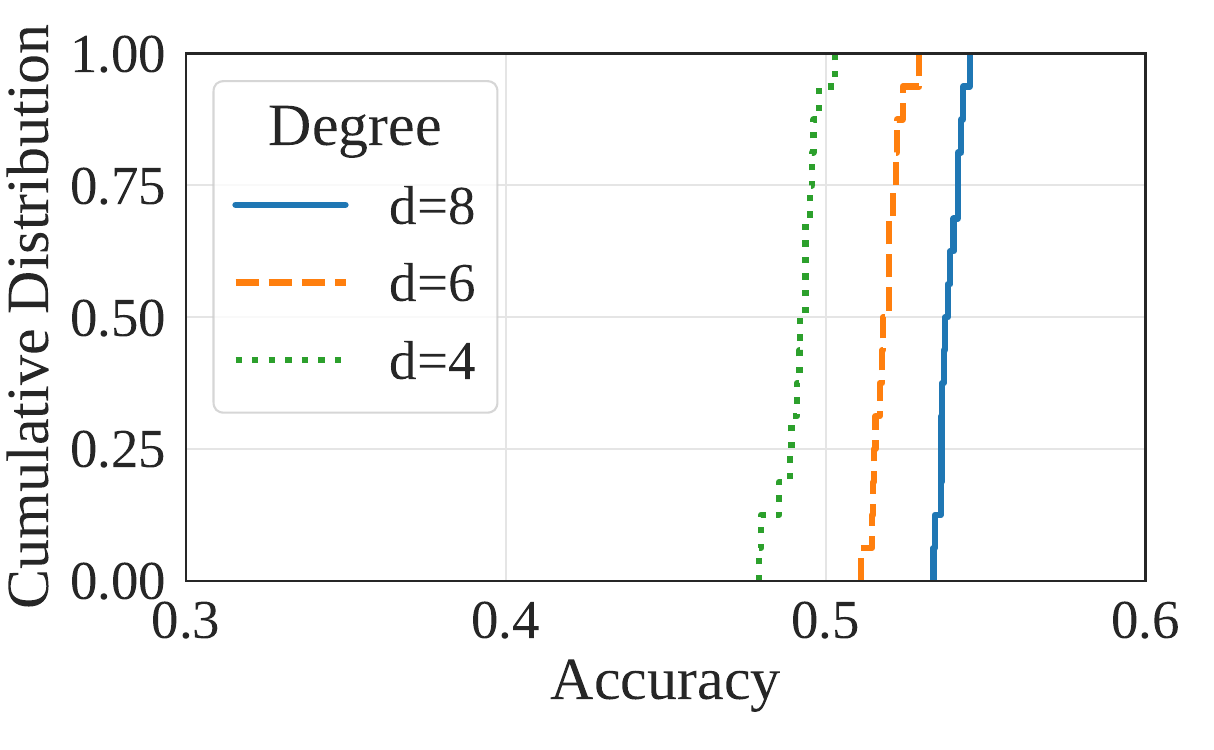}
        \vspace{-4.2ex}
        \caption{\footnotesize CIFAR-10. CDF of Accuracy}
        \label{fig:cdf-c-n16s8}
    \end{subfigure}
    \begin{subfigure}[b]{0.30\textwidth}
        \centering
        \includegraphics[width=\textwidth]{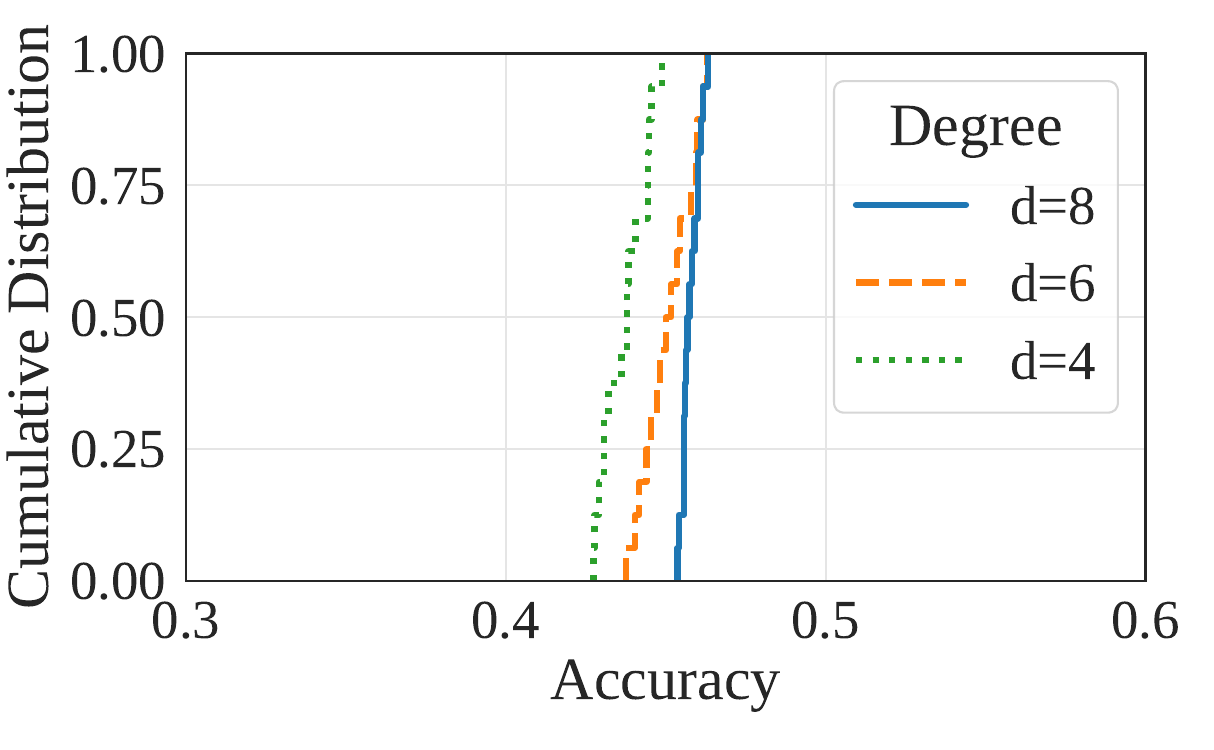}
        \vspace{-4.2ex}
        \caption{\footnotesize Shakespeare. CDF of Accuracy}
        \label{fig:cdf-s-n16}
    \end{subfigure}%
    \vspace{-1ex}
   \caption{Model accuracy from 16-client real experiments in Amazon EC2.} 
   \label{fig:acc-real}%
   \vspace{-3ex}
\end{figure*}

\subsubsection{ML datasets and models}
We evaluate the performance of FedLay for three ML tasks, including 1) Multilayer Perceptron (MLP) for digit classification on the MNIST dataset \cite{deng2012mnist}. 2) Convolutional Neural Networks (CNN) for image classification on the CIFAR-10 dataset \cite{cifar_09}. 3) Long Short-Term Memory (LSTM) for role forecasting on the Shakespeare dataset \cite{caldas2019leaf} built from \textit{The Complete Works of William Shakespeare}. Details are described in Table \ref{tab:expset}. All three are standard datasets for  FL benchmarks \cite{caldas2019leaf}.

\textbf{Learning with non-iid data.} We generate non-iid MNIST and CIFAR-10 datasets by selecting limited labels for the local training sets using the sharding method. Each shard contains only one label, and each local dataset includes a limited number of shards, resulting in a non-iid distribution and heterogeneity among clients' local datasets. 
For large-scale simulations, data among clients may overlap, but in real experiments and medium-scale simulations, clients have unique data. 
In the original Shakespeare dataset, each speaking role in a play is considered a unique shard.

\textbf{Client heterogeneity.} 
We also assumed that clients have different computation and communication resources. 
We set 3 tiers of clients. For the 16-client real-world experiments, we set 10 medium-capacity, 3 high-capacity, and 3 low-capacity clients. For simulations, each experiment includes 60\%  medium-, 20\% high-, and 20\%  low-capacity clients. The training time and communication time period of a high-capacity client are 2/3 of those of a medium-capacity user, and those of a low-capacity client are 2x of those of a medium-capacity user. 
The communication period for medium-capacity clients of each dataset is shown in Table \ref{tab:expset}.

\subsubsection{Performance metrics}
Besides the topology metrics discussed in Sec.~\ref{sec:metrics}, we use the following metrics to evaluate FedLay and other methods. 

\textbf{Model accuracy:} We evaluate the individual accuracy and average accuracy of local ML models based on separate test datasets that are different from the training datasets. 

\textbf{Topology correctness:} 
It is defined as the number of correct neighbors of all nodes over the total number of neighbors. Hence correctness equal to 1 means a correct FedLay.  


\textbf{Communication Cost:} We evaluate the average communication cost per client, by counting the number of NDMP messages sent by each client and the total size of the models sent by each client in bytes. 

\subsubsection{Methods for comparison}
There is no existing DFL topology that allows decentralized construction and maintenance. Hence we compare FedLay with the following methods:
1) Gaia \cite{gaia} is an ML method for geo-distributed clouds and still uses central servers. Hence it is not DFL. It runs server-based ML in each region and lets servers from different regions connect as a complete graph. It includes no aggregation method to handle non-iid data.   
2) DFL-DDS \cite{su2022boost} is a DFL method without a fixed topology. Instead, it simulates mobile nodes in a road network and considers two geographically close nodes as neighbors.  
3) Chord \cite{chord}. 
4) FedAvg \cite{pmlr-v54-mcmahan17a} is a standard centralized FL method. We use its accuracy as the \textit{upper bound} of DFL model accuracy because the central server knows all models from the clients.  


\begin{figure*}
    \centering
    \begin{subfigure}[b]{0.30\textwidth}
        \centering
        \includegraphics[width=\textwidth]{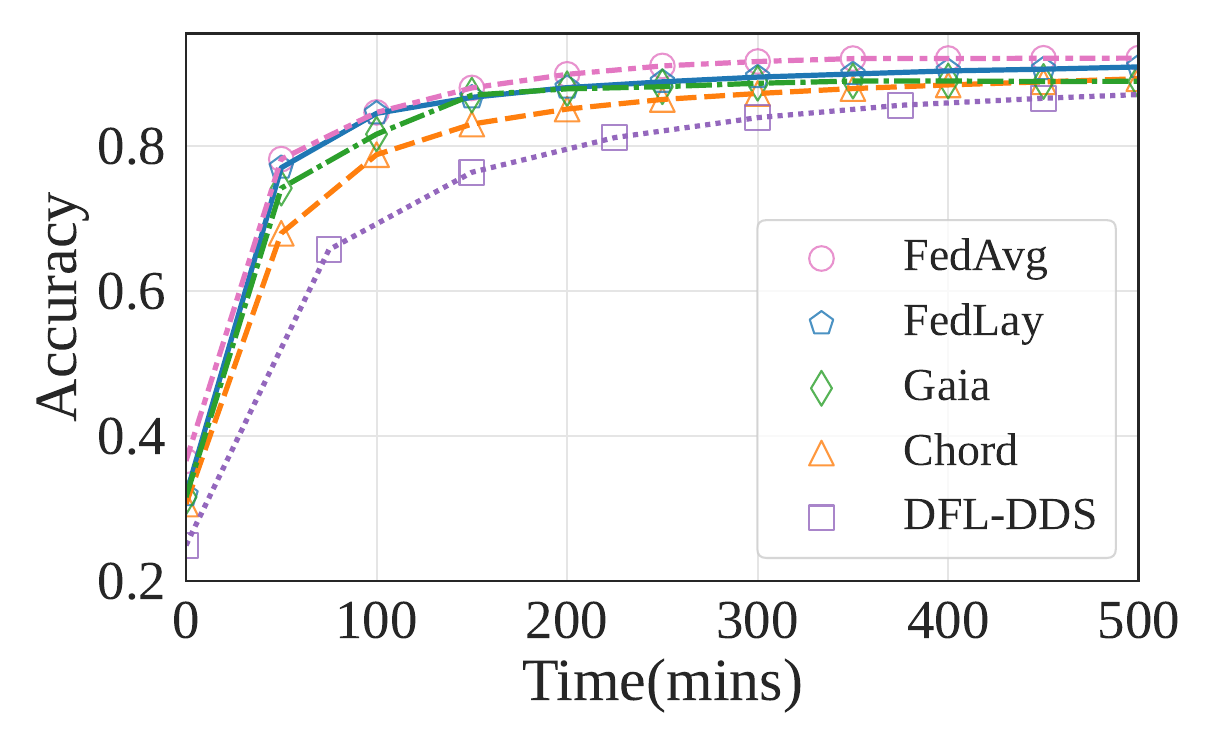}
        \vspace{-4.2ex}
        \caption{\footnotesize MNIST. Accuracy vs. Time}
        \label{fig:acc-m-n100}
    \end{subfigure}
    \begin{subfigure}[b]{0.30\textwidth}
        \centering
        \includegraphics[width=\textwidth]{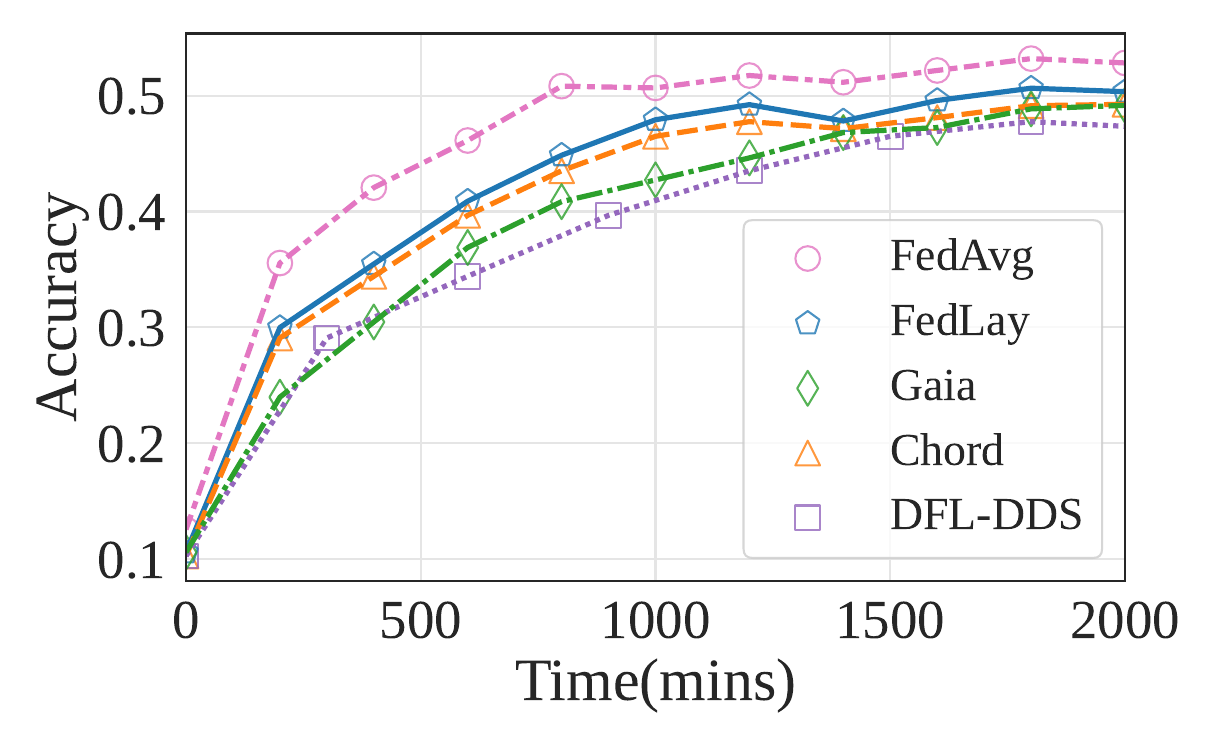}
        \vspace{-4.2ex}
        \caption{\footnotesize CIFAR-10. Accuracy vs. Time}
        \label{fig:acc-c-n100}
    \end{subfigure}
    \begin{subfigure}[b]{0.30\textwidth}
        \centering
        \includegraphics[width=\textwidth]{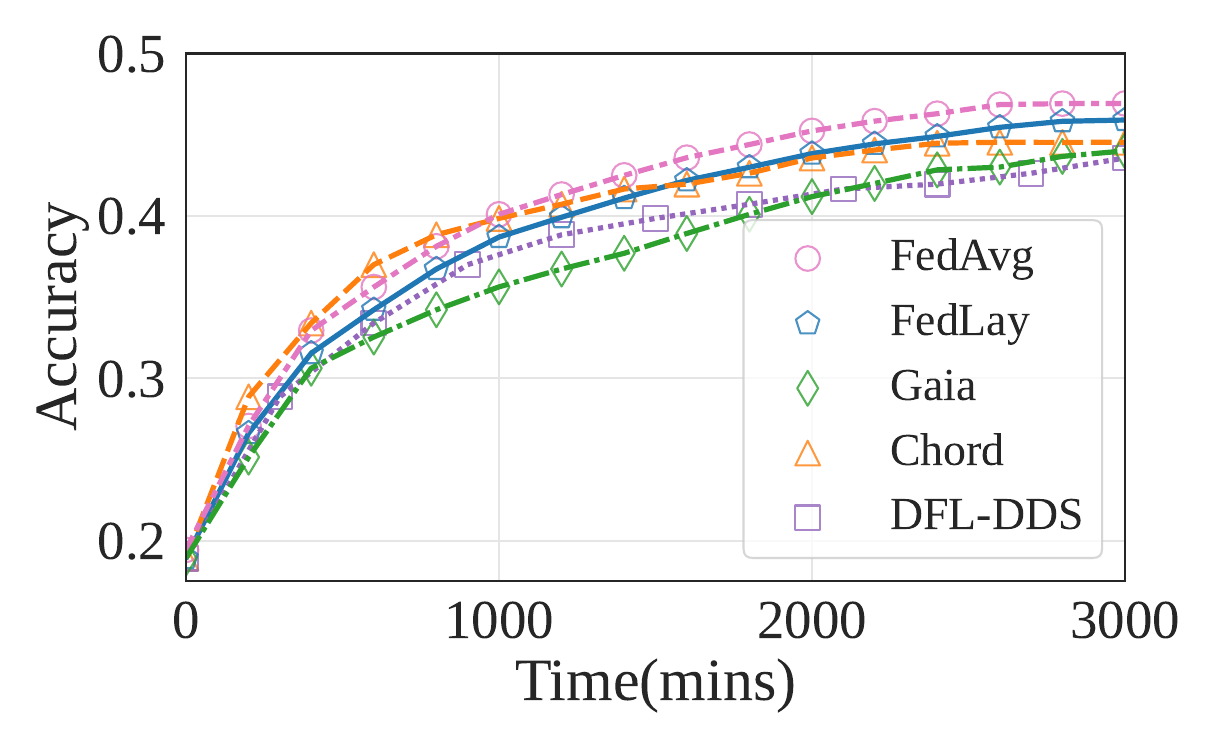}
        \vspace{-4.2ex}
        \caption{\footnotesize Shakespeare. Accuracy vs. Time}
        \label{fig:acc-s-n100}
    \end{subfigure}
    \vspace{-1ex}
   \caption{Model accuracy from 100-client medium-scale experiments. FedAvg is centralized and considered the upper bound. }
   \label{fig:acc-sim}
   \vspace{-2ex}
\end{figure*}

\begin{figure*}
    \centering
    \begin{subfigure}[b]{0.30\textwidth}
        \centering
        \includegraphics[width=\textwidth]{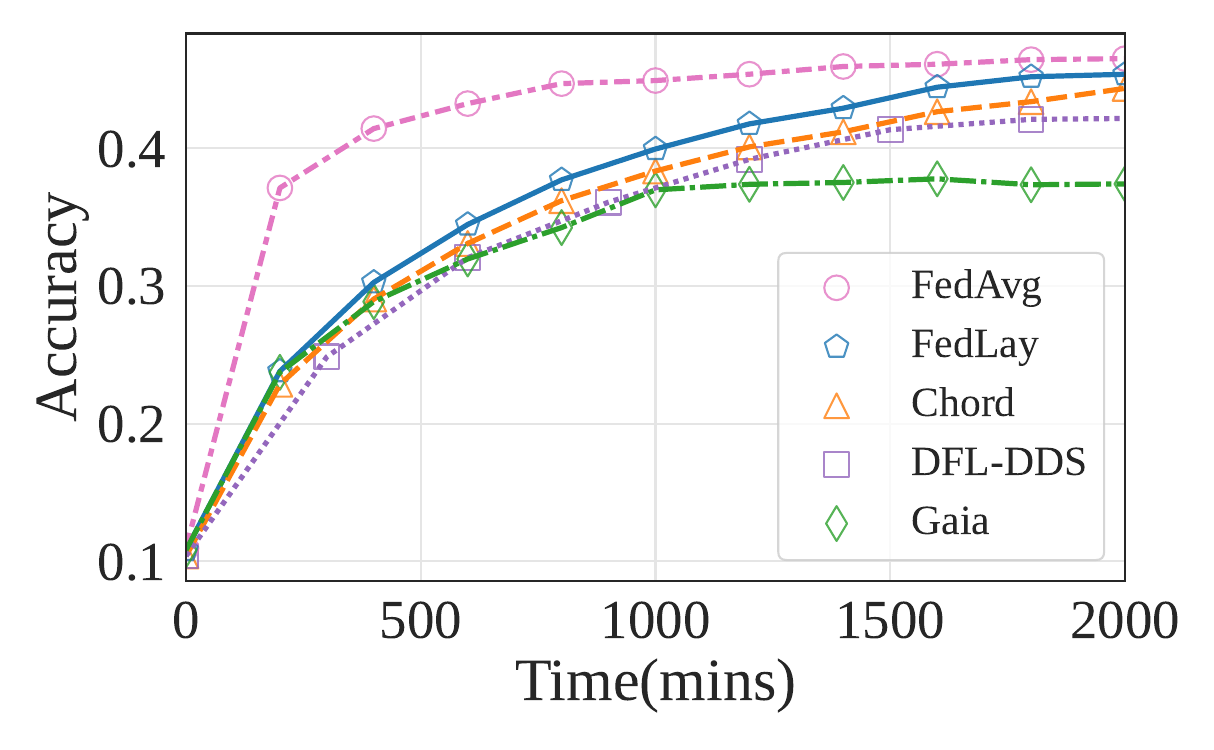}
        \vspace{-4.2ex}
        \caption{\footnotesize 4 shards per client}
        \label{fig:acc-c-n100s4}
    \end{subfigure}
    \begin{subfigure}[b]{0.30\textwidth}
        \centering
        \includegraphics[width=\textwidth]{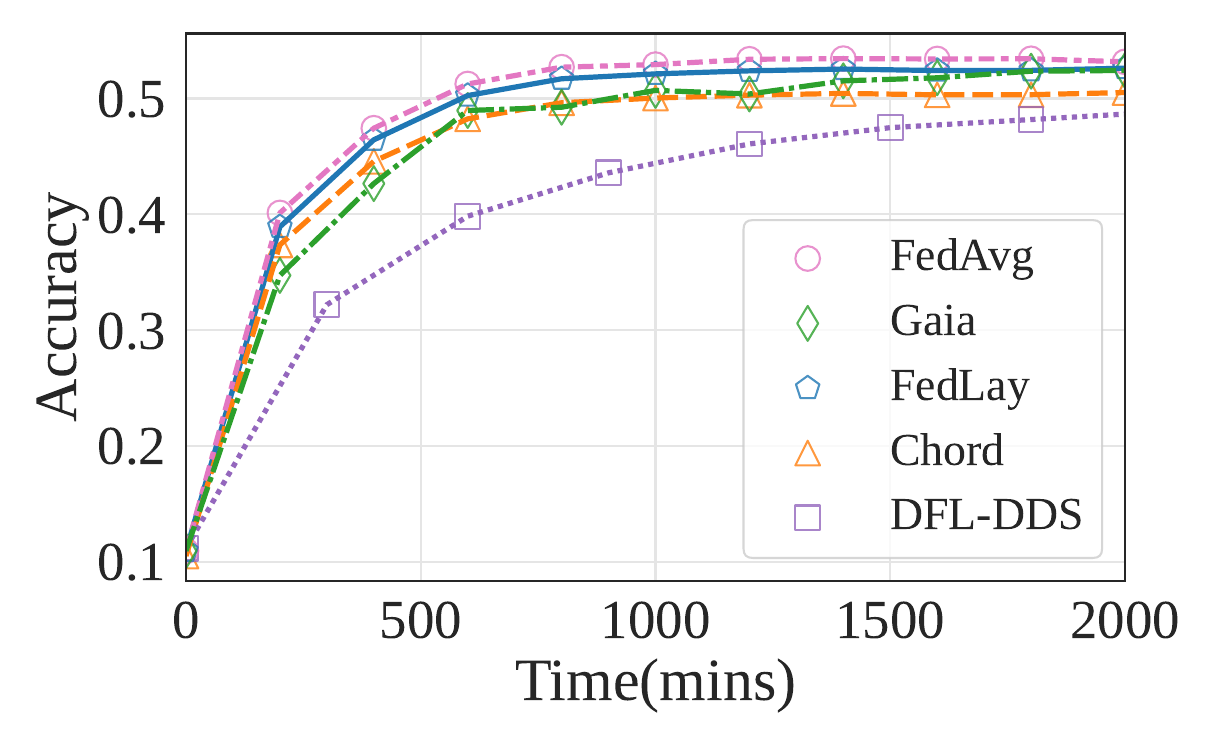}
       \vspace{-4.2ex}
        \caption{\footnotesize 12 shards per client}
        \label{fig:acc-c-n100s12}
    \end{subfigure}
    \begin{subfigure}[b]{0.30\textwidth}
        \centering
        \includegraphics[width=\textwidth]{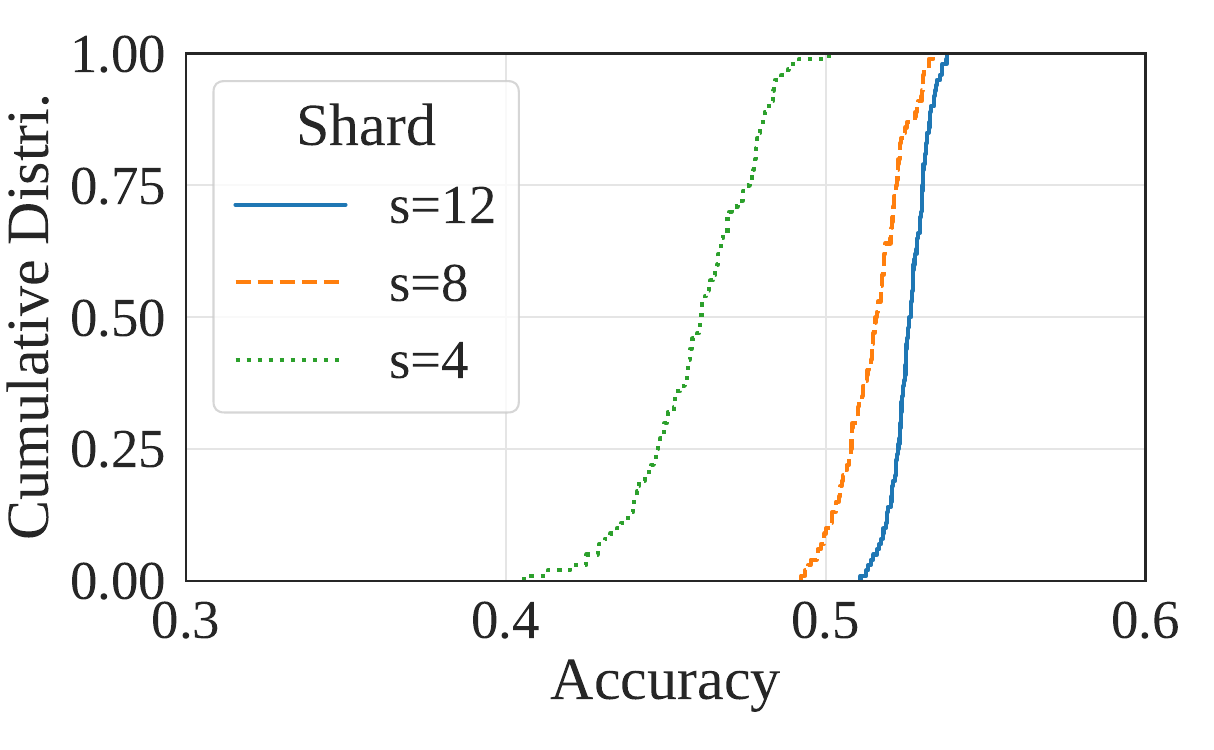}
        \vspace{-4.2ex}
        \caption{\footnotesize Accuracy distribution}
        \label{fig:cdfniid-c-n100}
    \end{subfigure}
    \vspace{-1ex}
   \caption{Accuracy under different non-iid levels for CIFAR-10.}
   \label{fig:acc-niid}
   \vspace{-2ex}
\end{figure*}

\vspace{-1ex}

\subsection{Evaluation of FedLay topology}
\vspace{-1ex}


We show the convergence factor, diameter, and average shortest path length by varying the number of nodes in Fig.~\ref{fig:expander3} for different topologies discussed in  Sec.~\ref{sec:fedlaytopology}. We evaluate FedLay with degrees of 6, 8, and 10, as well as Viceroy, Waxman, and Chord. 
We find that the diameter and average shortest path length of Viceroy and Waxman increases with the size of the network, and Chord has a large convergence factor when the number of nodes is large. It showed FedLay has the best results in all topology metrics.

Fig.~\ref{fig:expander4} shows the topology correctness under an extreme situation when 100 new clients join  a 400-client FedLay at the same time (10ms in the timeline). The average network latency is set to 350ms. 
We find the correctness can quickly converge to 1 after 8 seconds in FedLay with degree $d= 6, 8, 10, 12$. 
Fig. \ref{fig:expander5} shows the topology correctness under another extreme situation when 100 clients failed from a 400-client FedLay network at the same time (10ms in the timeline). 
The correctness quickly drops to 64.3\%. The remaining clients run NDMP and quickly recover to a correct 300-client FedLay network in 8 seconds.  
In Fig.~\ref{fig:expander6}, we plot the number of messages sent per client to construct FedLay networks with different sizes. With as many as 500 clients, each client only sends around 30 messages on average. 

\vspace{-1ex}
\subsection{DFL model accuracy}
\vspace{-1ex}
Fig.~\ref{fig:acc-real} shows the model accuracy of different methods in 6 subfigures, based on real experiments in Amazon EC2. 
In Figs.~\ref{fig:acc-m-n16s3}-\ref{fig:acc-s-n16} we find that FedLey achieves higher accuracy and faster convergence than Gaia and DFL-DDS, even with $d=4$.  
 Note one communication period for medium-capacity clients is set to 40 minutes in Shakespeare hence there are not many times of  model exchanges before convergence.  
Figs.~\ref{fig:cdf-m-n16s3}-\ref{fig:cdf-s-n16}  show the cumulative distribution of the accuracy of all clients at convergence time  (150 minutes for MNIST and 1500 minutes for others).
We can see that nodes are with similar accuracy levels without any `stragglers'. 

We evaluate the accuracy of FedLay ($d=10$), FedAvg, Gaia, and DFL-DDS using medium-scale experiments with 100 clients and show the results in Fig.~\ref{fig:acc-sim}. 
FedAvg achieves the 
best accuracy as a centralized FL, which we consider as the upper bound for DFL.
The accuracy of FedLay with 10 degrees is only 1.2\%, 2.5\%, and 0.9\% lower than FedAvg on MNIST, CIFAR-10, and Shakespeare, respectively.
Other methods have lower accuracy but the differences are not significant. 
We 
list accuracy at convergence time in Table \ref{tab:accuracy} \RE{, along with the default centralized method FedAvg as the baseline}.


\begin{table}[t]
    \fontsize{9}{9}\selectfont
    \centering
    \begin{tabular}{|c|c|c|c|c|c|}
        \hline
        \textbf{Task}  & \textbf{FedLay} & \textbf{FedAvg} & \textbf{Gaia} & \textbf{Chord} & \textbf{DFLDDS} \\
        \hline
        MNIST & 90.2\% & 92.1\% & 89.2\% & 88.9\% & 87.4\% \\
        \hline
        CIFAR & 50.3\% & 52.8\% & 48.6\% &  49.2\%& 49.4\%\\
        \hline
        Shakes & 45.9\% & 46.9\% & 44.0\% &44.5\% & 44.2\%\\
        \hline
    \end{tabular}
    \caption{Accuracy comparison at convergence. \RE{We regard the accuracy of FedAvg as the centralized baseline.}}
    \vspace{-4ex}
    \label{tab:accuracy}
\end{table}

We evaluate FedLay under different non-iid levels. Each client has a limited number of shards. When each client has fewer shards, the level of non-iid becomes more significant.
The default setting is 8 shards per client as shown in previous results such as  Fig.~\ref{fig:acc-c-n100},
Fig.~\ref{fig:acc-c-n100s4} and Fig.~\ref{fig:acc-c-n100s12} show the accuracy comparison for 4 shards per client and 12 shards per client respectively for CIFAR-10.  
We find that all DFL methods have slower convergence under more non-iid data (Fig.~\ref{fig:acc-c-n100s4}) but eventually FedLay still achieves similar accuracy as FedAvg, while Gaia and DFL-DDS have lower accuracy. 
We also show the distribution of accuracy of all clients at the time 2000 in Fig.~\ref{fig:cdfniid-c-n100}. 
When there are 4 shards per client, the distribution is more uneven. 
\begin{figure*}
    \centering
    \begin{subfigure}[b]{0.30\textwidth}
        \centering
        \includegraphics[width=\textwidth]{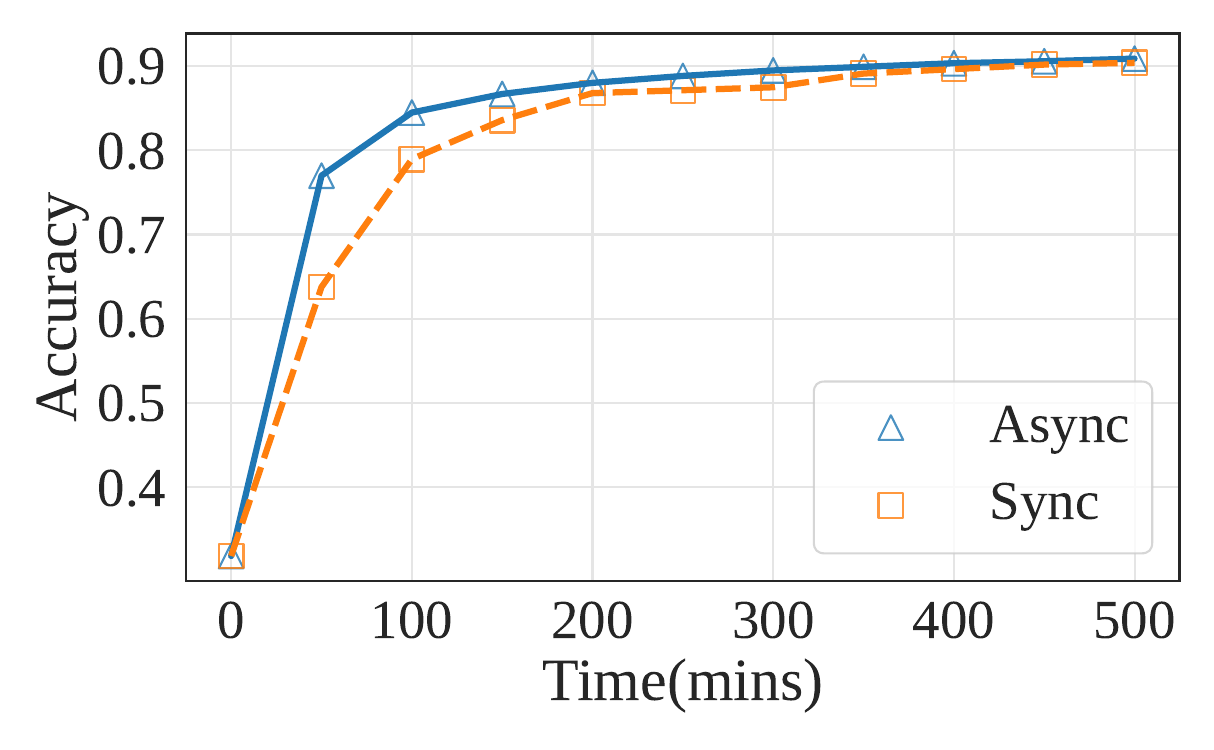}%
        \vspace{-2ex}
        \caption{\small  Accuracy for MNIST.}
        \label{fig:acc-m-n100-sync}
    \end{subfigure}
    \begin{subfigure}[b]{0.30\textwidth}
        \centering
        \includegraphics[width=\textwidth]{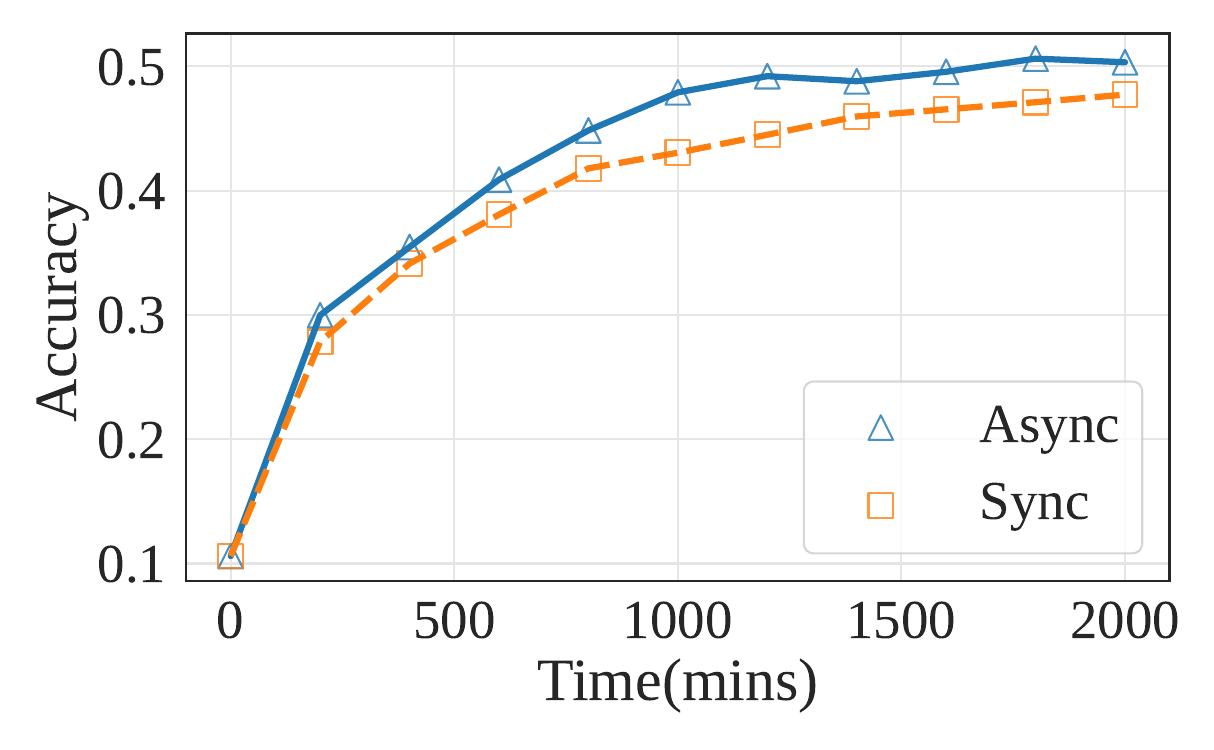}%
        \vspace{-2ex}
        \caption{\small  Accuracy for CIFAR-10.}
        \label{fig:acc-c-n100-sync}
    \end{subfigure}
    \begin{subfigure}[b]{0.30\textwidth}
        \centering
        \includegraphics[width=\textwidth]{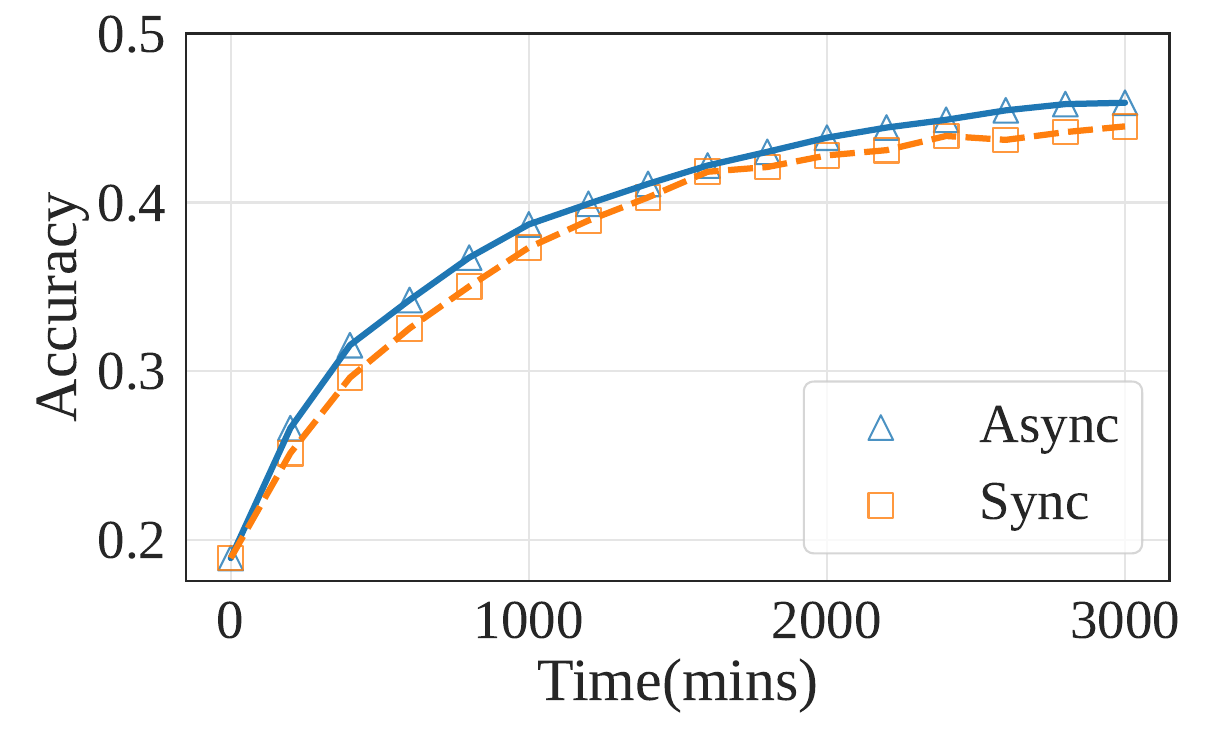}%
        \vspace{-2ex}
        \caption{\small  Accuracy for Shakespeare}
        \label{fig:acc-s-n100-sync}
    \end{subfigure}
    \vspace{-1ex}
   \caption{Model accuracy with synchronous and asynchronous communication.}
   \label{fig:acc-sync}
   \vspace{-3ex}
\end{figure*}

\begin{figure*}[t!]
\centering
\begin{tabular}{p{160pt}p{160pt}p{160pt}}
	\centering
        \includegraphics[width=0.23\textwidth]{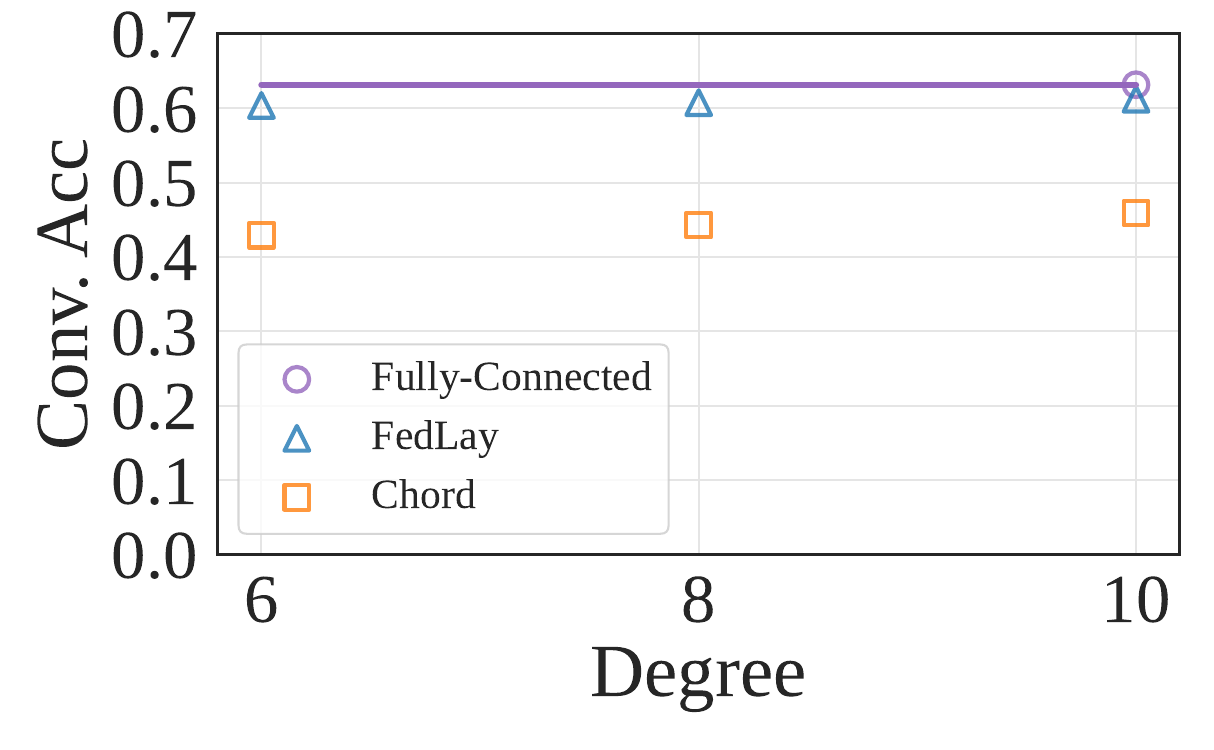}
	\vspace{-2.5ex}
	\caption{Converged Accuracy}
	\label{fig:locality-degree}
	&
       \hspace{-2ex}
	\centering\includegraphics[width=0.24\textwidth]{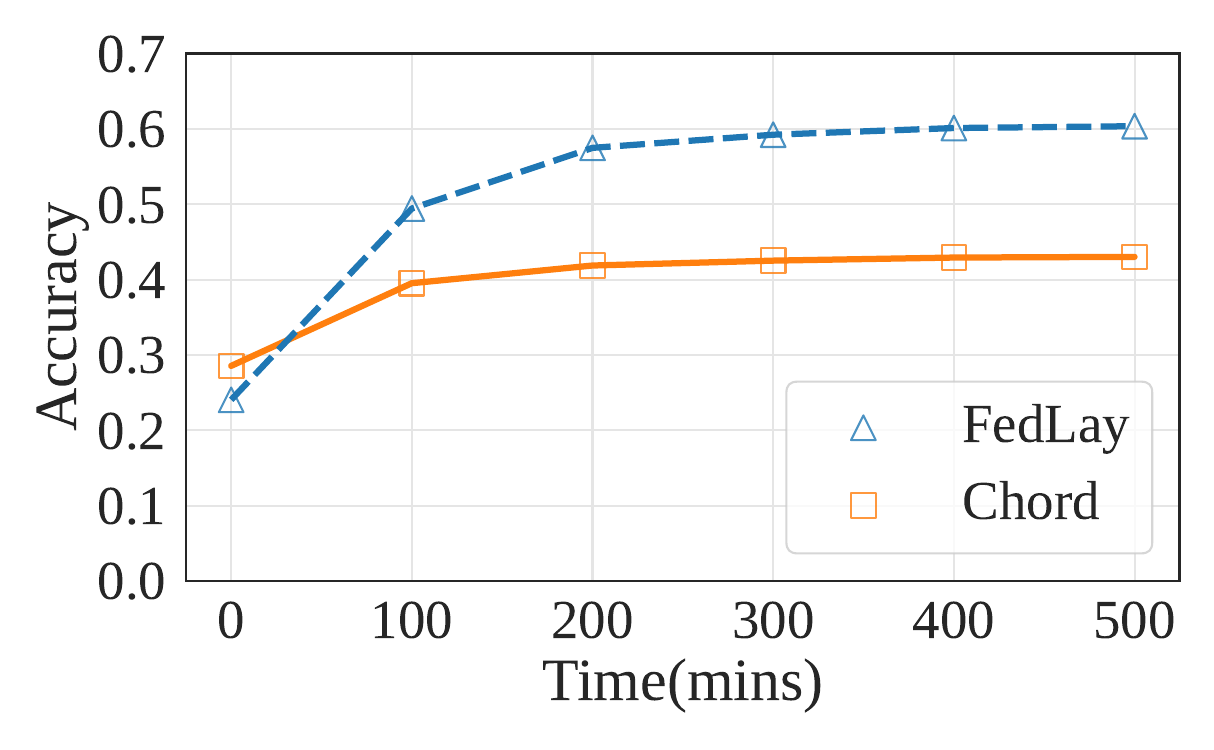}
	\vspace{-1.5ex}
	\caption{Accuracy vs Time(d = 8)}
        \label{fig:locality-time}
	&
       \hspace{-5ex}
	\centering\includegraphics[width=0.24\textwidth]{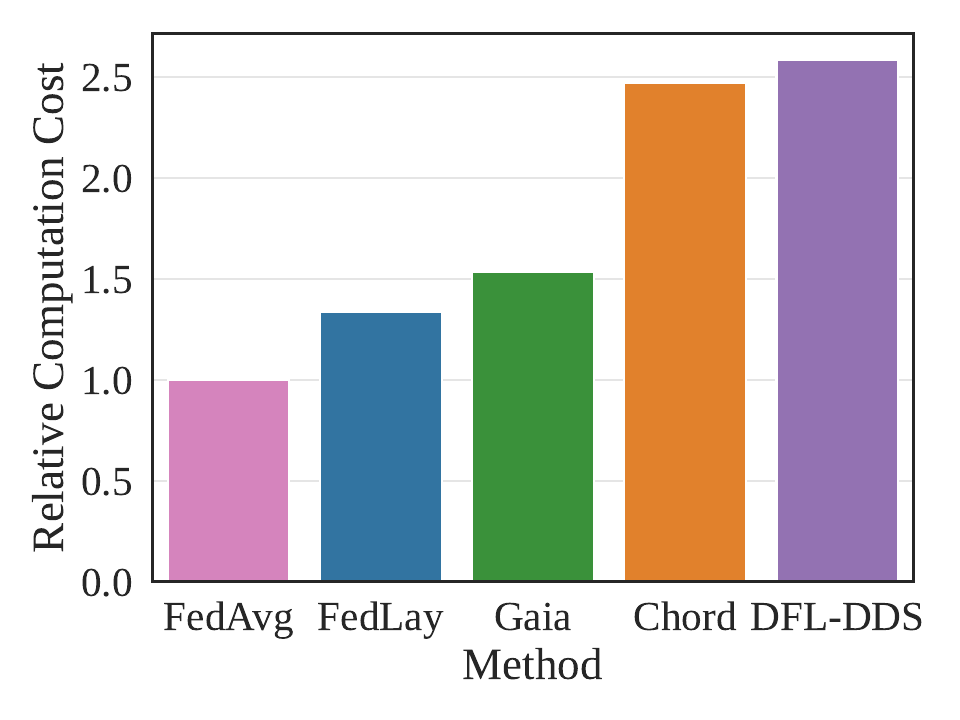}
	\vspace{-1.5ex}
	\caption{Relative Computational Cost}
        \label{fig:rcc-m-n100s3}
\end{tabular}
\vspace{-3ex}
\end{figure*}

\textbf{Evaluation of data with biased distribution and locality.}
In this set of experiments, 
100 clients are divided into 10 groups evenly, and each group possesses 6 out of the total 10 labels in the CIFAR-10 dataset. Each group only has 1 label that is different from the neighboring groups. For example, group 1 has labels 1 to 6; group 2 has labels 2 to 7, etc. and the last group has labels 10, 1, 2, 3, 4, 5. For each client, we sampled 2000 images for each label evenly from the original CIFAR-10 dataset. In Fig.\ref{fig:locality-degree}, we show FedLay has an average of $37.01\%$ improvement over Chord on varying degrees. It also demonstrates FedLay is only $2.0\%$ lower than the theoretical upper bound, a fully connected network. In Fig.\ref{fig:locality-time}, we show the comparison of the accuracy of FedLay and Chord versus time. Again, FedLay shows much better convergence compared to Chord. 

\vspace{-1ex}
\subsection{Evaluation of other considerations}
\vspace{-1ex}

\textbf{Asynchronous communication.} 
We also compare FedLay with synchronous and asynchronous communication in Fig.~\ref{fig:acc-sync}. 
We find for all three datasets, asynchronous communication can improve both the accuracy and convergence speed, because high-capacity clients do not need to wait for low-capacity ones. 

\begin{figure}[t!]
\centering

\vspace{-2ex}
\begin{tabular}{p{120pt}p{120pt}}
	\centering\includegraphics[width=0.23\textwidth]{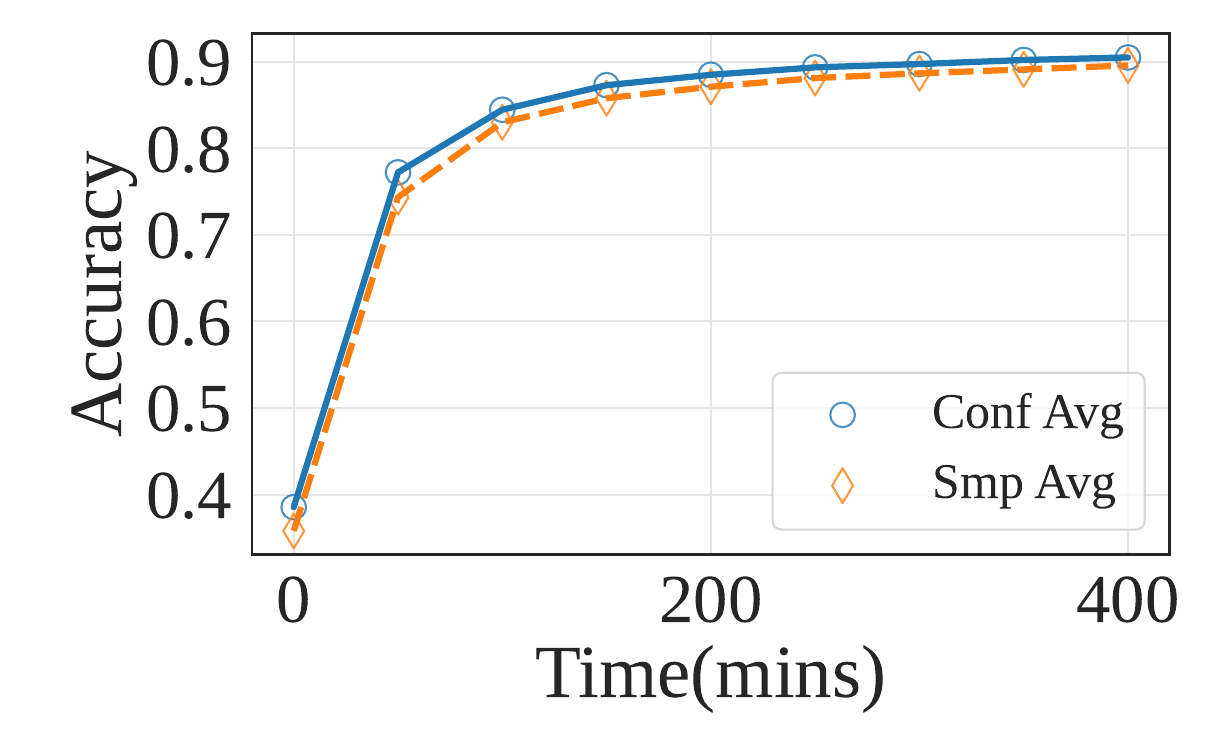}
	\vspace{-4.5ex}
	\caption{Accuracy for MNIST, $\alpha_d = 0.5,\alpha_c = 0.5$ }
        \vspace{-2ex}
	\label{fig:confacc-m-n100d10s3-1}
	&
       \hspace{-5ex}
	\centering\includegraphics[width=0.24\textwidth]{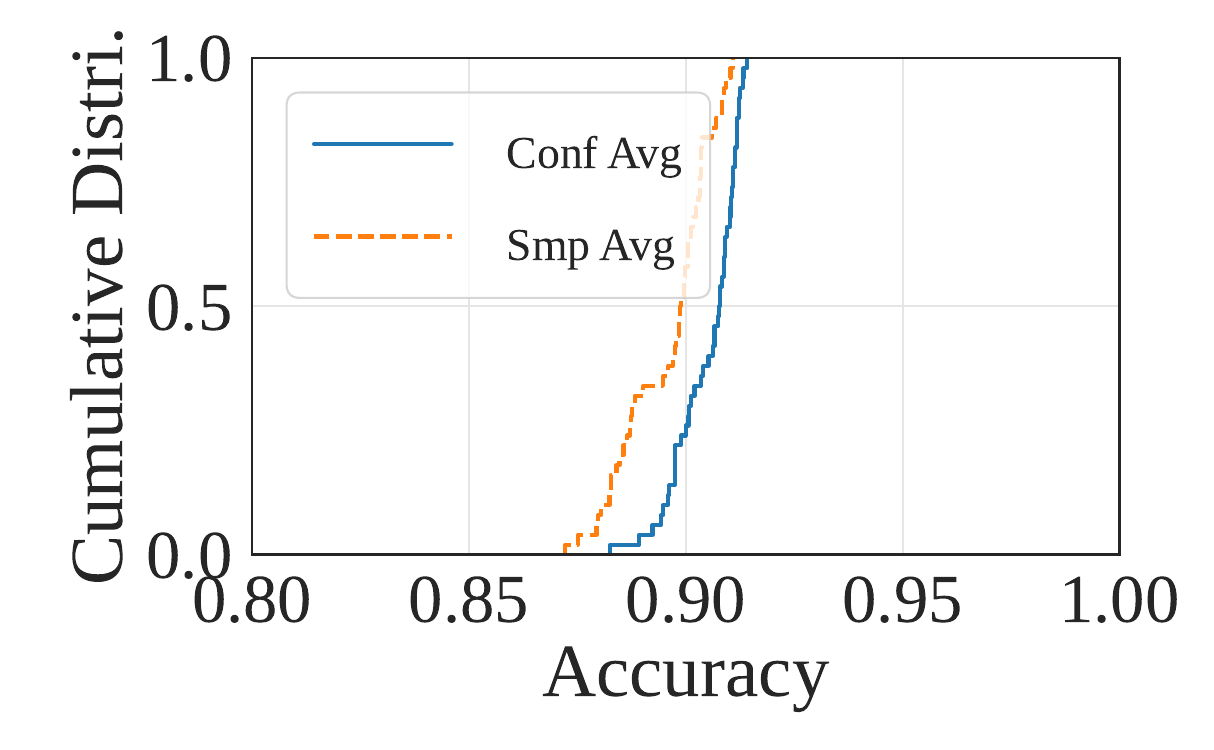}
	\vspace{-1.5ex}
	\caption{Accuracy distribution for MNIST}
        \vspace{-2ex}
        \label{fig:confcdf-m-n100d10s3-1}
\end{tabular}
\vspace{-5ex}
\begin{tabular}{p{120pt}p{120pt}}
	\centering\includegraphics[width=0.23\textwidth]{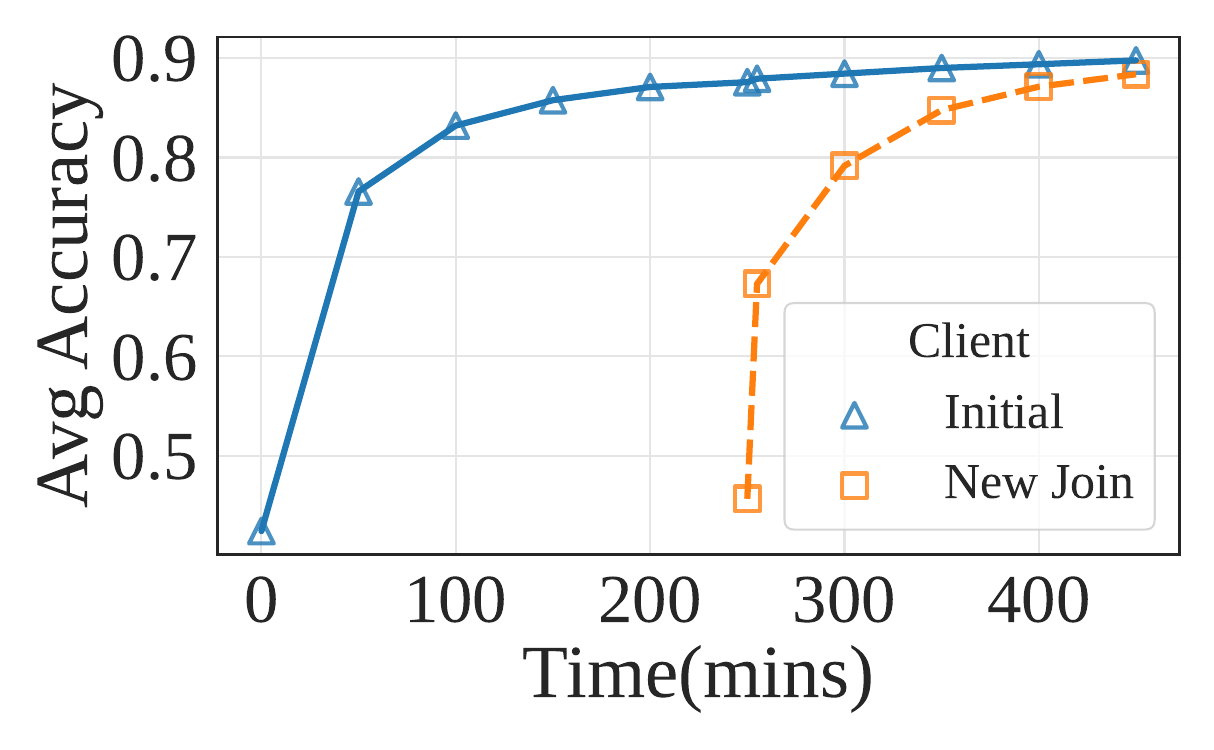}
	\vspace{-4.5ex}
	\caption{Accuracy under churn.}
        \vspace{-2.5ex}
        \label{fig:join-m-n5050d10s4-1}
	&
 \hspace{-5ex}
	\centering\includegraphics[width=0.23\textwidth]{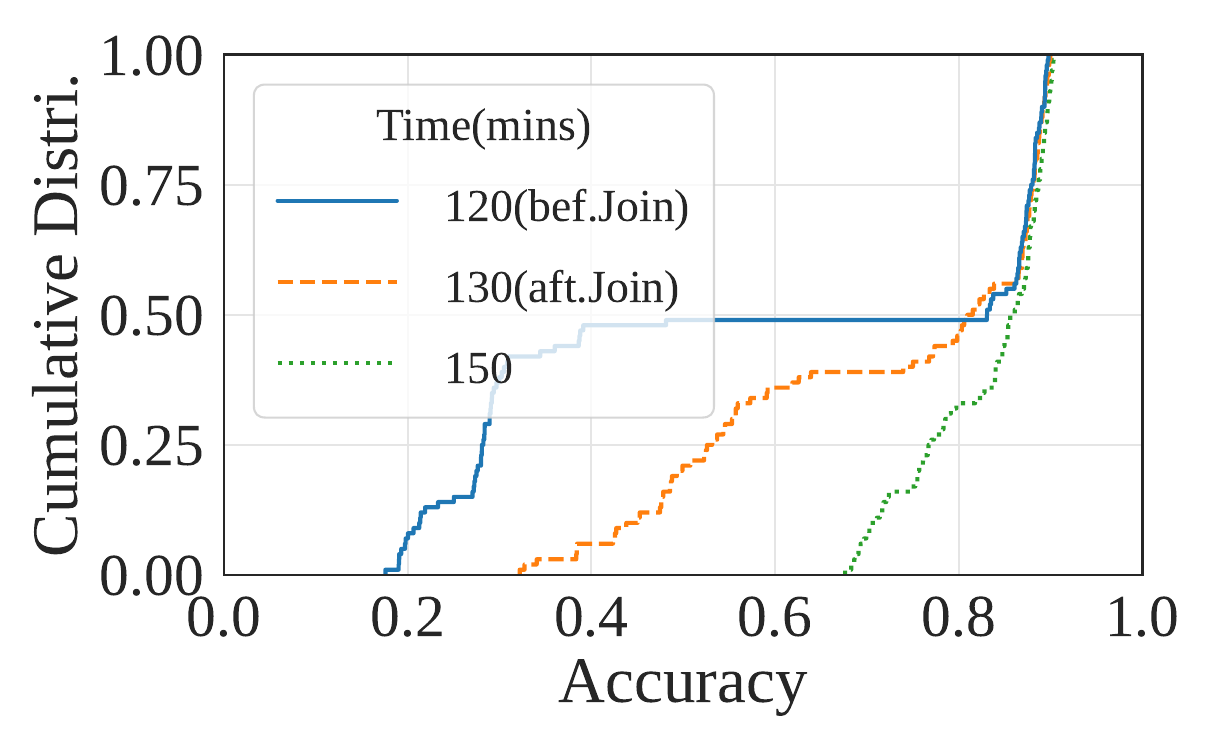}
	\vspace{-1.5ex}
	\caption{Accuracy distribution under churn.}
        \label{fig:joincdf-m-n5050d10s4-1}
        \vspace{-2.5ex}
\end{tabular}
\vspace{-2.5ex}
\end{figure}

\textbf{Confidence parameters.} 
Fig.~\ref{fig:confacc-m-n100d10s3-1} and Fig.~\ref{fig:confcdf-m-n100d10s3-1} shows the accuracy of FedLay with and without confidence parameters for MNIST, compared to simple average. 
We set $\alpha_d=0.5$ and $\alpha_c=0.5$. The results show that FedLay slightly improves the simple average in accuracy. 

\textbf{Accuracy under churn.} 
We show the model accuracy under extreme churn: 50 new clients join a 50-client FedLay network.
In Fig.~\ref{fig:join-m-n5050d10s4-1}, the curves with triangle markers show the accuracy of the initial 50 nodes and the curves with square markers show the accuracy of 50 newly joined nodes. We find that the accuracy of the new nodes quickly converges to a high level due to the high-confidence models from existing nodes. 
Fig.~\ref{fig:joincdf-m-n5050d10s4-1} show that at the join time, the newly joined nodes have very low accuracy and all clients achieve high accuracy eventually.  

\RE{\textbf{Computation Cost}. We show the relative computational cost in Fig.\ref{fig:rcc-m-n100s3}. In the experiment of 100 nodes training on MNSIT dataset, To reach the accuracy of 88\%, The relative computation cost of FedLay is 1.33, compared to 1.53 for Gaia, 2.47 for Chord, and 2.76 for DFL-DDS, with the baseline FedAvg normalized to 1. FedLay only has 33\% overheads, smallest compared to other methods.}


\vspace{-2ex}
\subsection{Scalability}
\vspace{-1ex}

\begin{figure}[t]
    \centering
    \begin{subfigure}[b]{0.23\textwidth}
        \centering
        \includegraphics[width=\textwidth]{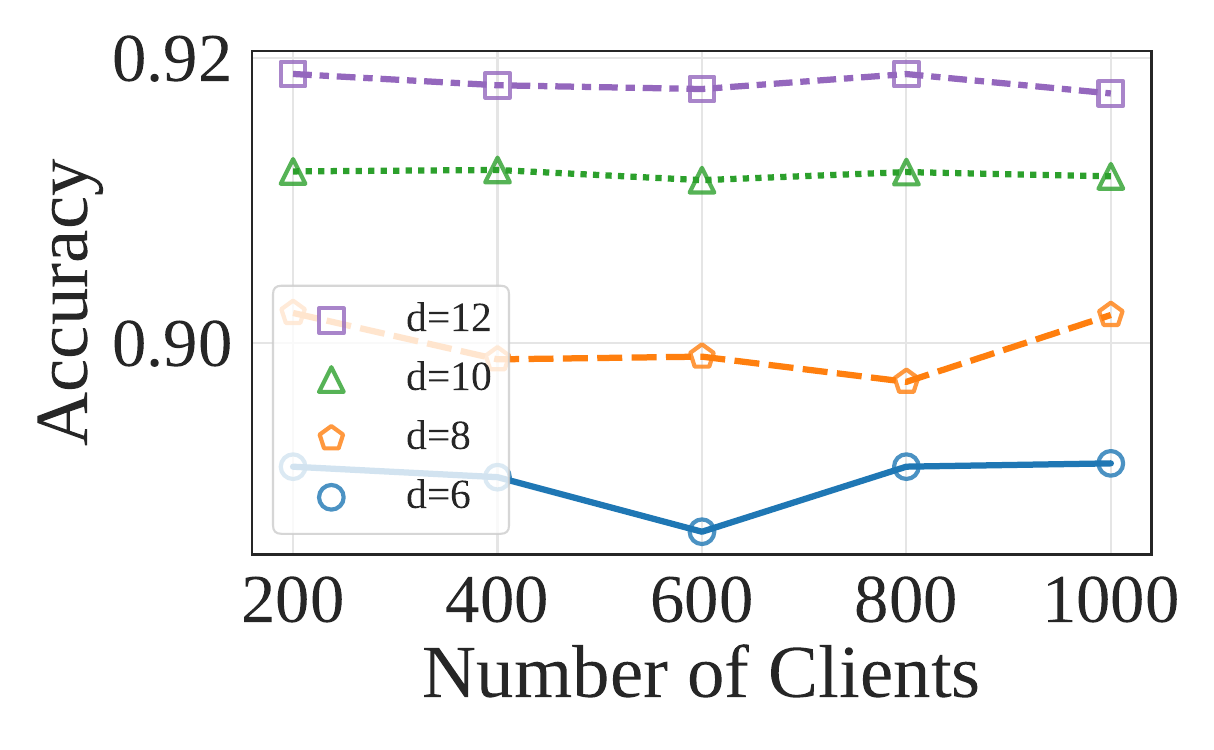}
        \vspace{-2ex}
        \caption{\footnotesize MNIST }
        \label{fig:acc-m-nlarge}
    \end{subfigure}
    \begin{subfigure}[b]{0.225\textwidth}
        \centering
        \includegraphics[width=\textwidth]{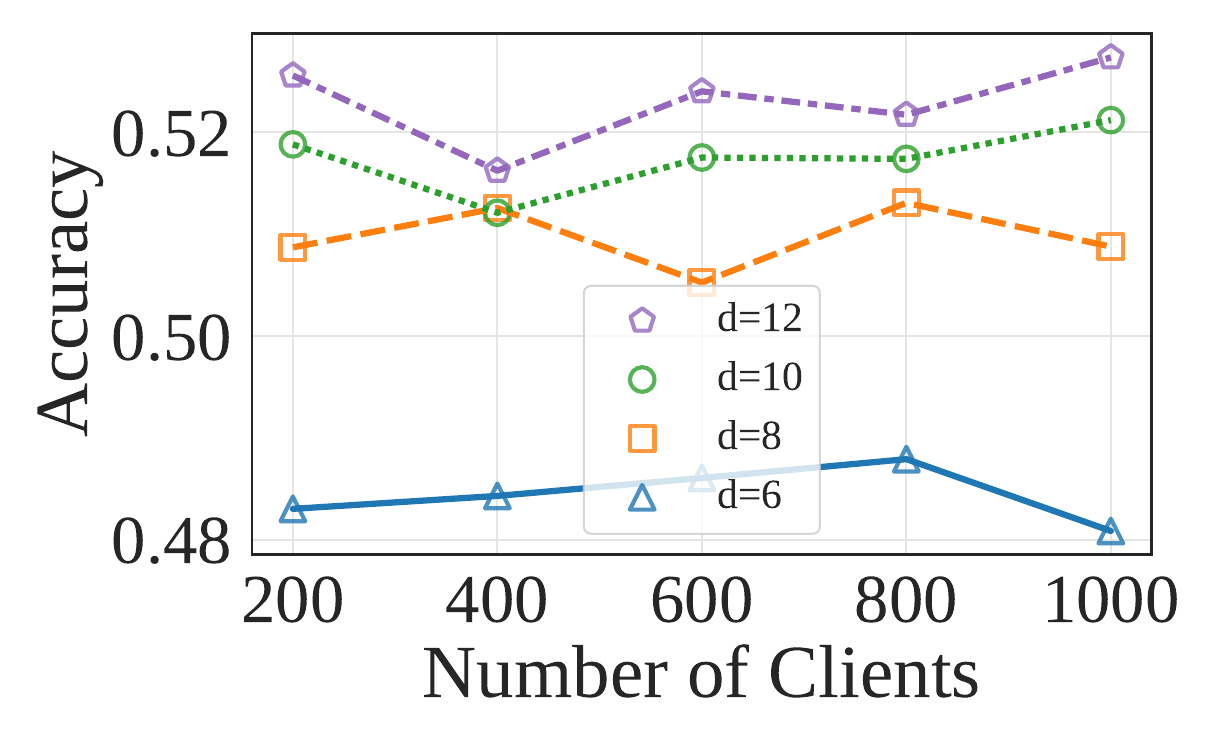}
        \vspace{-2ex}
        \caption{\footnotesize CIFAR-10 }
        \label{fig:acc-c-nlarge}
    \end{subfigure}
    \begin{subfigure}[b]{0.23\textwidth}
        \centering
        \includegraphics[width=\textwidth]{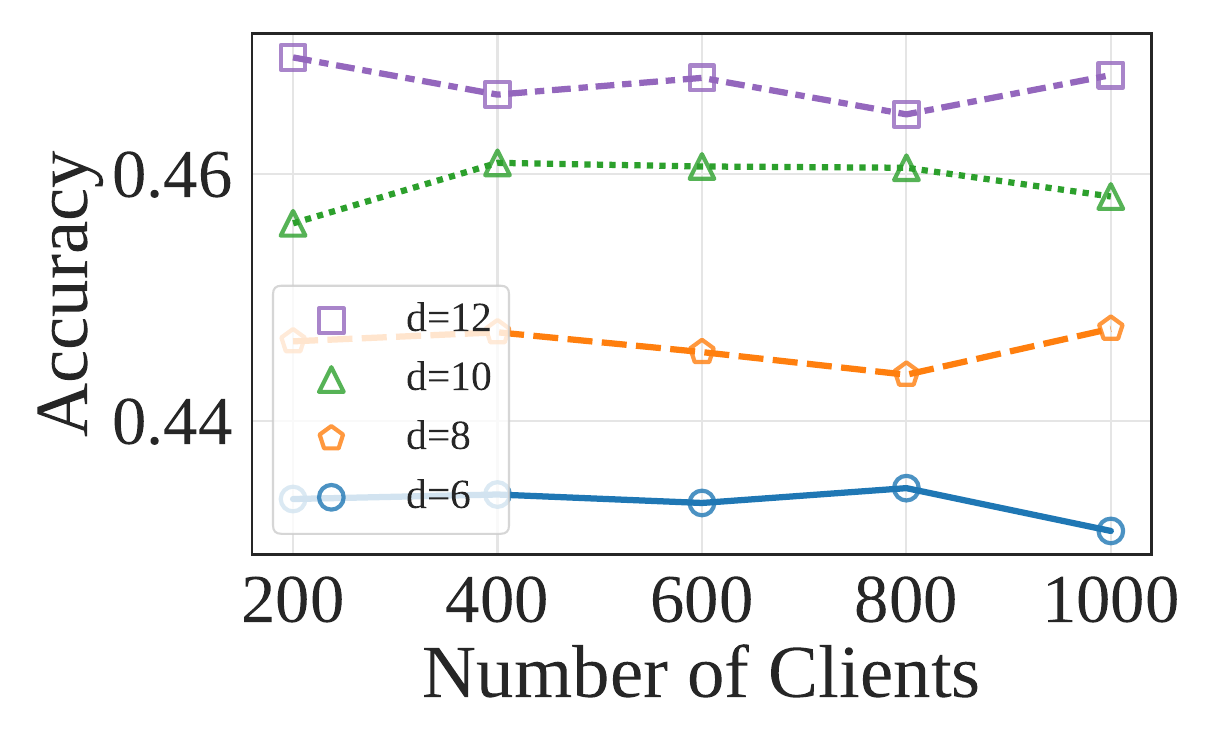}
        \vspace{-2ex}
        \caption{\footnotesize Shakespeare }
        \label{fig:acc-s-nlarge}
    \end{subfigure}
    \begin{subfigure}[b]{0.24\textwidth}
        \centering
        \includegraphics[width=\textwidth]{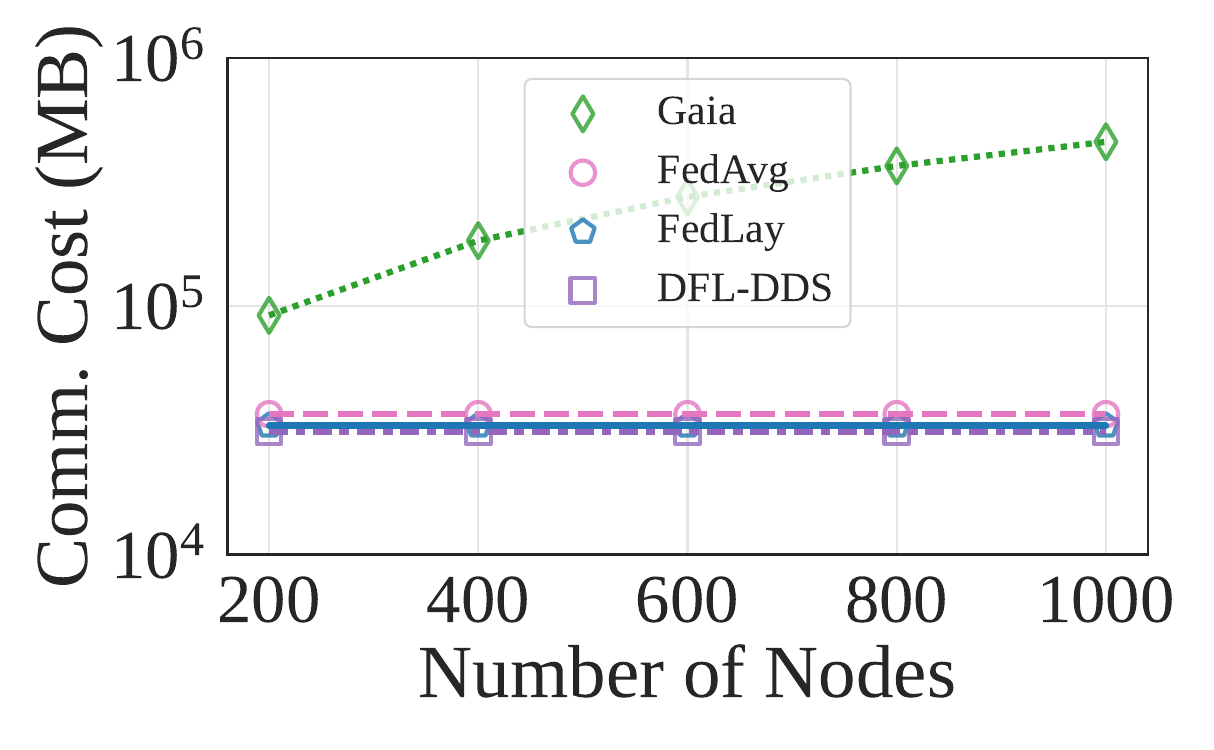}
        \vspace{-5ex}
        \caption{\footnotesize Shakespeare: Comm. cost}
        \label{fig:comm-nlarge}
    \end{subfigure}
    \vspace{-1ex}
   \caption{Results of simulations for large-scale networks}
   \label{fig:acc-emu}
   \vspace{-4.2ex}
\end{figure}

We use large-scale simulations to evaluate the scalability of  FedLay, as shown in Fig. \ref{fig:acc-c-nlarge}. We find that even with up to 1000 clients, FedLay has stable performance in all  datasets. In Fig.~\ref{fig:comm-nlarge}, we compared the communication cost  per client (in MBs) to reach the convergence of FedLay to those of FedAvg, DFL-DDS, and Gaia. 
Gaia has poor scalability in communication. 

\vspace{-1ex}
\section{Related Works}
\label{sec:related}
\vspace{-1ex}

\textbf{Decentralized Federated Learning (DFL).} Federated learning (FL) \cite{pmlr-v54-mcmahan17a} is an attractive solution for large-scale ML that allows many clients to train 
ML models collaboratively without directly sharing training data. However, the central server in  FL is a single point of failure and attack.   
Decentralized Federated Learning (DFL) has been proposed to remove the central server \cite{he2018cola,sun2021decentralized,Beltran2022DFL}. 
The overlay network of DFL is a fundamental problem. He \textit{et al.} \cite{he2018cola} suggest a few overlay  topologies including ring, 2D grid, and complete graph, which either are unable to be constructed in a decentralized way or cause too much communication. GADMM \cite{elgabli2020communication} uses a dynamic chain topology and other methods apply clustering-based topologies \cite{al-abiad2022cluster} \cite{bellet2022d}. 
Vogels \textit{et al.} \cite{vogels2022beyond} analyze model convergence in theory on  different topologies including hypercube, torus, binary tree, and ring. Recently Hua \textit{et al.}  suggest applying Ramanujan graphs for DFL \cite{hua2022efficient}. No above studies discuss decentralized construction and maintenance of the suggested topologies. \RE{Recently, \cite{zhang2023secure, xu2022spdl} suggest to utlize Blockchain enhance the security and verifiability of DFL}. 

\textbf{Other overlay topologies.} Overlay topologies have been extensively studied for P2P networks. A well-known category of overlay networks is called distributed hash tables (DHTs), such as Chord \cite{chord}.  
DHTs are proposed to achieve data searching in a decentralized network. Distributed Delaunay triangulation (DT) \cite{DT-ICNP,MDT} is designed to achieve greedy routing guarantees and Viceroy \cite{Viceroy} is designed for minimizing congestion by overlay routing. Near-random regular topologies have been studied for data center networks with centralized construction \cite{Jellyfish,S2-ICNP}.

\vspace{-1ex}
\section{Conclusion}
\label{sec:conclusion}

This work presents FedLay, the first overlay network for DFL that achieves all the following properties: 1) decentralized
construction, 2) fast convergence to accurate models, 3) small node degree, and 4) resilience to churn. 
We present the detailed designs of the FedLay topology, neighbor discovery and maintenance protocols (NDMP), and model exchange protocol (MEP). 
We prove that NDMP can guarantee the correctness of a decentralized overlay for node joins and failures. 
Evaluation results show that FedLay provides the highest model accuracy
among existing DFL methods, small communication costs,
and strong resilience to churn. In particular, it provides significant model accuracy advantages compared to other decentralized protocols such as Chord, when data are distributed with locality and bias. 

\clearpage
 \bibliographystyle{plain}
 \bibliography{references}

\end{document}